\documentclass[11pt,a4paper]{article}
\pdfoutput=1
\usepackage{jheppub}

\usepackage{hyperref}
\usepackage{amsfonts}
\usepackage{amsmath}
\usepackage{amssymb}
\usepackage{graphicx}
\usepackage{color}
\usepackage{textcomp}
\usepackage{bm}

\def\be{\begin{equation}}
\def\te{\end{equation}}
\def\ee{\end{equation}}
\def\ba{\begin{eqnarray}}
\def\bea{\begin{eqnarray}}
\def\nn{\nonumber\\}
\def\tea{\end{eqnarray}}
\def\ea{\end{eqnarray}}
\def\eea{\end{eqnarray}}

\title{Nonlinear Fluctuations in Relativistic Causal Fluids}

\author[a]{Nahuel Miron-Granese}
\emailAdd{nahuelmg@df.uba.ar}
\author[b]{Alejandra Kandus}
\emailAdd{kandus@uesc.br}
\author[c]{Esteban Calzetta}
\emailAdd{calzetta@df.uba.ar}

\affiliation[a]{Departamento de F\'isica, Facultad de Ciencias Exactas y Naturales, Universidad de Buenos Aires, Cuidad Universitaria, 
Buenos Aires 1428, Argentina.}
\affiliation[b]{Departamento de Ci\^encias Exatas e Tecnol\'ogicas, Universidade Estadual de Santa Cruz, Rodov. Jorge Amado km 16,
CEP: 45.662-900, Ilh\'eus - BA, Brasil.}
\affiliation[c]{Universidad de Buenos Aires, Facultad de Ciencias Exactas y Naturales. Departamento de F\'\i sica, Buenos
Aires C1428EGA, Argentina. \\
CONICET-Universidad de Buenos Aires, Facultad de Ciencias Exactas y Naturales, Instituto de F\'\i sica de Buenos Aires
(IFIBA), Buenos Aires C1428EGA, Argentina.}

\abstract{
In the Second Order Theories (SOT) of real relativistic fluids, the non-ideal properties of the flows are described by a new set
of dynamical tensor variables. In this work we explore the 
non-linear dynamics of those variables in a conformal fluid. 
Among all possible SOTs, we choose to work with the Divergence Type Theories (DTT) formalism, which ensures that the second 
law of thermodynamics is fulfilled non-perturbatively. The tensor modes include two divergence-free modes which have no analog 
in theories based on covariant generalizations of the Navier-Stokes equation, and that are particularly relevant because they
couple linearly to a gravitational field. To study the dynamics of this irreducible tensor sector, we observe that in causal
theories such as DTTs, thermal fluctuations induce a stochastic stirring force, which excites the tensor modes while preserving
energy momentum conservation. From fluctuation-dissipation considerations it follows that the random force is Gaussian with a white spectrum. 
The irreducible tensor modes in turn excite vector modes, which back-react on the tensor sector, thus producing a consistent 
non-linear, second order description of the divergence-free
tensor dynamics. Using the Martin-Siggia-Rose (MSR) formalism plus the Two-Particle Irreducible Effective Action (2PIEA)
formalism, we obtain the one-loop corrected
equations for the relevant two-point correlation functions of the model: the retarded propagator and
the Hadamard function. The overall result of the self-consistent dynamics of the irreducible tensor modes at this order is a depletion of the  spectrum in the UV sector, which suggests that tensor modes
could sustain an inverse entropy cascade.
}


\arxivnumber{2002.08323}

\begin{document}

\maketitle

\section{Introduction}\label{intro}

Fluid description of relativistic matter proved to be a powerful tool for a clearer understanding of high energy phenomena
\cite{landau1,landau2}. Examples are the thermalization \cite{romat17} and isotropization \cite{Stri14} of the quark-gluon plasma 
created in the Relativistic Heavy Ion Collider (RHIC) facilities, the behaviour of matter in the inner cores of Neutron Stars (NS) \cite{rishke10,FriedSterg13,sterg17} and during cosmological phase transitions \cite{nikschlesigl18}, etc. The processes observed in those systems cannot be explained using ideal relativistic fluids.

Unlike non-relativistic hydrodynamics, where there is a successful theory to describe non-ideal fluids, namely the Navier-Stokes 
equation, there is no definite mathematical model to study real relativistic fluids. The history of the development of such theory 
begins with the recognition of the parabolic character of Navier-Stokes and Fourier equations\footnote[1]{Recall that the 
non-relativistic Fourier law allows for an instantaneous propagation of heat.} \cite{israel88}, which implies that they cannot be 
naively extended to relativistic regimes. In fact, the first attempts by Eckart and Landau \cite{Eck40,LL6} to build a relativistic 
theory of dissipative fluids starting from the non-relativistic formulation, also encountered this pathology.

The paradox about the non-causal structure of Navier-Stokes and Fourier equations,  known as First Order Theories (FOTs), was
resolved phenomenologically in 1967 by I. M\"uller \cite{muller67}. He showed that by including second order terms in the heat flow 
and the stresses in the conventional expression for the entropy, it was possible to obtain a system of phenomenological equations 
which was consistent with the linearized form of Grad kinetic equations \cite{grad49}, i.e., equations that describe transient effects 
that propagate with finite velocities. These equations, constitute the so-called Second Order Theories (SOTs), whose main difference
with respect to FOTs is that the stresses are upgraded to dynamical variables that satisfy a set of Maxwell-Cattaneo equations
\cite{Max67,Catt48,Catt58,JosPrez89}, thus constituting a
hyperbolic theory.
Later on, M\"uller's phenomenological theory was extended to the relativistic regime by W. Israel and J. M. Stewart, and others 
\cite{israel76,IsSte76,IsSte79a,IsSte79b,IsSte80,OlsHis90,JouRubiCV80,PavJouCV80,HisLind85,HisLind88a,HisLind88b,Ols90,JouCVLe10,Noronha19}. 

The problem with parabolic evolution equations
is not restricted to relativistic fluids. Hyperbolic theories
are useful not only in the relativistic domain, but also in the
non-relativistic one, whenever the relaxation time toward a steady
state is larger than the time between collisions as, e.g., second sound in superfluids and solids, polymeric fluids, etc. \cite{AnPavRo98,HerrPav01}. The difference between parabolic and hyperbolic theories can be ignored only when the relaxation times are shorter than the characteristic time scale of the state. This was originally suggested by Maxwell \cite{Max67}.
Muronga \cite{Muronga04} compared the descriptions of early 
stages in heavy ion collisions given by ideal, FOTs and SOTs, 
and concluded that the latter were the most accurate and 
reliable among them.

The rationale for pursuing second order theories such as DTTs or anisotropic hydrodynamics \cite{Stri14} also comes from the observation that highly anisotropic expansions such as in the early stages of RHICs, produce strong local momentum anisotropies that cannot be described considering weak deviations of the distribution function from local equilibrium. Also, hydrodynamic formalisms derived from Grad's theory are liable to predict spurious instabilities not present in the underlying kinetic theory, for example, when considering perturbations of highly anisotropic flows \cite{milton17}, while SOTs give a more accurate description. As a matter of fact, not all SOTs are able to coup with highly anisotropic states: even for moderate specific shear viscosities $\eta/s \sim 5-10$ the (negative) longitudinal component of the viscous shear pressure can become so large in Israel-Stewart theory, that it overwhelms the thermal pressure, resulting in a negative total pressure along the beam direction \cite{GaJeSche13,JeHei15}.

In 1996 Liu, M\"uller and Ruggieri \cite{LiMuRu86} developed a field-like description of particle density, particle flux and energy-momentum components. The resulting 
field equations were the conservation of particle number, energy momentum and balance of fluxes, and were strongly constrained by the 
relativity principle, the requirement of hyperbolicity and the entropy principle. The only unknown functions of the formalism were the 
shear and bulk viscosities and the heat conductivity, and all propagation speeds were finite. Several years later, Geroch and Lindblom extended the analysis of Liu et al. and wrote down a general theory were all the dynamical equations can be written as total-divergence equations \cite{GerLind90,GerLind91}, see also Refs. \cite{cal98,ReNa97,PRCal09,PRCal10,cal15,milton17,LheReRu18,lucas19}. This theory, known as Divergence 
Type Theory (DTT) is causal in an open set of states around equilibrium states. Moreover, all the dynamics is determined by a single scalar 
generating functional of the dynamical variables, a fact that allows to cast the theory in a simple mathematical form. Besides the dynamical 
equations an extra four-vector current is introduced, the entropy four-current, which is a function of the basic fields and not of any of their derivatives, and
whose divergence is non-negative by the sole virtue of 
the dynamical equations. This fact guarantees that the second law is 
automatically enforced at all orders in a perturbative development. In contrast, as Israel-Stewart-like theories must be built order by 
order, the Second Law of thermodynamics must be enforced in each step of the construction \cite{Loga08}. In other words, DTTs are exact 
hydrodynamic theories that do not rely on velocity gradient expansions and therefore go beyond Israel-Stewart-like second-order theories. Further, DTTs have been tested through its application to Bjorken and Gubser flows \cite{lucas19}, where exact solutions of the kinetic theory are available.

The novelty of SOTs, is the introduction of tensor dynamical variables to account for non-ideal features of the flow. This means that 
besides the scalar (spin 0) and vector (spin 1) modes already present in Landau-Lifshitz or Eckart theories, it is possible to excite tensor 
(spin 2) perturbations. This fact enlarges the set of hydrodynamic effects that a real relativistic fluid can sustain. 
As is familiar for the gravitational field, the tensor sector can be further decomposed  into scalar, vector, and divergence-free components. 
If present in the Early Universe plasma, the latter could excite primordial gravitational waves \cite{nahuel17}, or seed primordial 
electromagnetic fluctuations \cite{calkan16}. Another scenario where tensor modes could play a relevant role are high energy astrophysical 
compact objects as, e.g., Neutron Stars \cite{FriedSterg13}. It is well known that rotational tensor normal modes of those stars can source gravitational waves, however at present there is no compelling hydrodynamical model for them.

The study of hydrodynamic fluctuations in the relativistic regime is a natural follow-up to the successful application of the theory to describe thermalization and isotropization in systems such as RHICs \cite{romat17}. Among all the possible lines to pursue 
this task, there is the study of the stochastic flows induced by the fluid's own thermal fluctuations.

The development of fluctuating hydrodynamics was pioneered by Landau and Lifshitz \cite{LLflucthyd,LL6,LLsmII,FoxUhl1,FoxUhl2}, who applied
the fluctuation dissipation theorem \cite{CallWel} to the Navier-Stokes equation. Hydrodynamic fluctuations in a consistent relativistic theory were studied in \cite{JouRubiCV80} and, in the context of DTTs, in \cite{cal98}. Forcing by thermal noise is relevant to the calculation of the transport coefficients of the fluid \cite{MS11,KMS11,Jero15} as well as the 
phenomenon of \emph{long time tails} \cite{AlWain70,HauLee70,KY03,KMS12,Teaney17,Martinez18,ABSY19}. 
Noise, whether thermal or not, can also play an important 
role in early Universe phenomena such as primordial magnetic field induction \cite{kacamawa00,calkan02} and phase transitions 
\cite{calver99,calrouver01} to cite a few.

In the literature on relativistic fluctuating hydrodynamics, no distinctions were made between scalar and vector fluctuations, 
not even tensor fluctuations were mentioned. So, as the distinctive feature of SOTs is that they sustain tensor modes, in this work 
we shall concentrate on the study of fluctuations in the pure tensor sector. Among all possible SOTs, we choose to work with DTTs. 
The reasons were mentioned above: they are thermodynamically and relativistically consistent in arbitrary flows and independently 
of any approximations. Consequently in a perturbative development we do not have to worry about enforcing the Second Law.

To address the subjet of study, we use effective field theory methods, which long ago began to be used in the study of turbulence
\cite{Hopf52,LewKrai62,DeDomMar79}, and continue to be a powerful tool to study random flows \cite{Polonyi15,CGL15,JPY18,HLR18}. 
Among all methods, the Effective Action formalism allows to express the different N-point correlation functions of the theory
in terms of loop diagrams, which adds a new source of intuition in the intepretation of the correlations. In this manuscript we
use the Two-Particle Irreducible Effective Action (2PIEA) formalism, through which we write down the evolution equations for the 
relevant two-point functions of the problem under study, namely the Retarded and the Hadamard propagators. This field method 
allows to build the propagators as the contribution of all closed interacting diagrams that cannot be separated by cutting two of 
their internal lines. 

The Martin-Siggia-Rose (MSR) formalism \cite{msr-73,dedo-76,kam-11,eyink96,zancal02,mcomb14} is a systematic way to derive the sought 2PIEA, which then may be straightforwardly evaluated through a diagrammatic expansion \cite{wyld61,lee65,mcomb90,mcomb14,cal09}. 
These methods have been applied to relativistic fluids in Refs. \cite{romat17,Kovtun12,KMR14,HKR15}. To characterize the thermal
fluctuations we shall use a formulation of the fluctuation - dissipation theorem appropriate to causal hydrodynamics \cite{cal98}, 
recovering the Landau-Lifshitz fluctuating hydrodynamics as a limiting case.

In summary, the relativistic perfect fluid approach that is widely used in several high energy descriptions
is rather well established \cite{AndCom06}, while relativistic
viscous hydrodynamics is much less well understood. We are
dealing with systems whose fundamental description, whether
kinetic theory, field theory or a combination of both, involves
many more degrees of freedom than hydrodynamics may capture. As most excitations decay exponentially fast, long lived modes can be considered an attractor in the space of solutions \cite{HelSpal15,romatschke17,StricNorDen18,DenNor19} regardless of the initial conditions. In this sense stochastic SOTs represent phenomenological attempts to model the longest-lived non-hydrodynamic modes as additional degrees of freedom (besides the 
usual hydrodynamic ones) and the short-lived as noise, in timescales 
of the order or shorter than the relaxation time, using causality as a guideline. Among all SOTs, DTTs have the advantage of being thermodynamically consistent. It is important to mention 
that recently there appeared several works indicating that causality 
might be satisfied within FOTs 
\cite{VanBiro12,kovtun19,DasFlorNoRy20} (see also \cite{GPRuRe19}). 
However our interest in SOTs is precisely the fact that its range of variables is broader, including specially divergenceless tensor modes which enable a richer description of irreversible processes. Further, it is well-known that the transverse and traceless projection of the anisotropic stress tensor, $\Pi_{\rm TT}^{ij}$, plays an important role in the dynamics of the gravitational waves acting as a source \cite{weinberg2008}. In the cosmological context there are several scenarios in which matter has non-vanishing $\Pi_{\rm TT}^{ij}$ and it affects the amplitude of gravitational waves \cite{capriniandfigueroa}, e.g. freely streaming neutrinos or photons, 
interacting scalar fields and primordial density perturbations, among others.

The paper is organized as follows. 
In section \ref{gdtt} we begin by quickly reviewing the Landau-Lifshitz hydrodynamics, after
which we build the minimal conformal Divergence-Type-Theory beyond LL. We end this section by setting the criterion for 
incompressibility.
In section \ref{MSREA} we give an abridged presentation of the fluctuation-dissipation theorem in DTTs consistent with a causal 
theory \cite{cal98} and outline the 
Martin-Siggia-Rose formalism for the Two-Particle-Irreducible Effective Action. We briefly show how this formalism allows to write 
down evolution equations for the main propagators of the theory, namely the retarded (or causal) propagator and the Hadamard two-point 
function.
In Section \ref{LFAE} we study the effect of linear fluctuations around an equilibrium state and find the lowest order causal
propagators as well as the vector and tensor Hadamard propagators. In Section \ref{Nlfae} we extended the analysis of Section \ref{LFAE} to include nonlinear fluctuations. We work within the free-streaming
approximation, which amounts to consider times
shorter than the typical macroscopic relaxation time. Even at these very early times one may formulate a consistent hydrodynamics that acts as an attractor for the evolution of the system \cite{romatschke17,StricNorDen18,DenNor19}. The free-streaming approximation is equivalent to the 
approximation $\tau \to \infty$ which  was recently shown to be appropriate for the early stages in 
high-energy heavy ion collisions \cite{KMPOS20}.
In this regime we find that at one-loop in the interactions and using dimensional regularization 
to treat the ultraviolet divergences, the equation 
for the causal tensor propagator acquires a new term which renormalizes the relaxation time introducing a scale dependence proportional to the fourth power of the momentum. In consequence the Hadamard correlation function is also modified in a way that shows 
a transition from a flat spectrum, for low values of $p$, to a power law spectrum ($\propto p^{-4}$) with increasing $p$.

Before going on a comment is in order. In the large-$\tau$ limit we are considering there may be several so-called non-hydrodynamical modes contributing to the dynamics. We have chosen to include only a single tensor mode as representative, because such a mode is naturally present in the energy-momentum tensor, and so it is easily identifiable in a large range of models. However, in the absence of a symmetry principle that singles out this tensor mode above other possible slow modes, such as higher spin currents, these extra modes may contribute similarly to loop corrections. In such a case, a more complex theory must be considered and the analysis presented in this paper is a preliminary first step in this direction. We end Section \ref{Nlfae} by briefly discussing the effect of the tensor fluctuations on the mean value of the entropy, which is also depleted in the ``large-$p$ range'' of the spectrum, arguing that this may be an indication of an inverse cascade of entropy \cite{Eyink18b}. 
Finally in Section \ref{concl} we discuss the results we obtained, draw our main conclusions and suggest possible lines to pursue the study developed in this manuscript. We left for the Appendix \ref{app:loopscaling} the discussion of the
scaling law of the main diagrams that contribute to the one-loop
approximation.

\section{The model}\label{gdtt}

To make this manuscript self-contained, we begin this section with a condensed review of Landau-Lifshitz hydrodynamics for a conformal
neutral fluid, in order to show that a FOT does not guarantee fulfillment of the Second Law of thermodynamics.
Among all possible SOTs, we choose a DTT to build what is arguably the minimal extension of Landau-Lifshitz hydrodynamics which enforces 
the Second Law of thermodynamics non-perturbatively, and where the dynamics of the neutral fluid is given by the conservation laws of the 
energy-momentum tensor (EMT) $T^{\mu\nu}$ and of a third order tensor $A^{\mu\nu\rho}$ that encodes the non-ideal properties of the flow. 
The theory is completed by considering an entropy current $S^{\mu}$ whose conservation equation enforces the Second Law of 
thermodynamics. $T^{\mu\nu}$ is symmetric and traceless, and $A^{\mu\nu\rho}$ is totally symmetric and traceless on any two indices
\cite{cal013}. We linearize the evolution equations and find the propagation speed for the scalar, vector and tensor modes, from which 
we write down a criterion to define incompressibility.

\subsection{Landau-Lifshitz hydrodynamics in a nut-shell}\label{gdtt-LL}

Let us consider the simplest model for a conformal fluid, for which there is no particle number current and the energy-momentum tensor 
is traceless. The energy density $\rho$ is defined by the Landau prescription
\begin{equation}
T^{\mu\nu}u_{\nu} = -\rho u^{\mu}\label{apa-1}
\end{equation}
with normalization $u^2 = -1$. Observe that eq. (\ref{apa-1}) is also the definition of $u^{\mu}$ as an eigenvector of $T^{\mu\nu}$
with eigenvalue $-\rho$. For an ideal fluid the energy momentum tensor must be isotropic in the rest frame, so
\begin{equation}
T^{\mu\nu}_0 = \rho u^{\mu}u^{\nu} + p \Delta^{\mu\nu}\label{apa-2}
\end{equation}
where 
\be
\Delta^{\mu\nu} = g^{\mu\nu} + u^{\mu}u^{\nu} \label{delta}
\te
is the projector onto hypersurfaces orthogonal to $u^{\mu}$. Tracelessness implies the equation of state
\begin{equation}
p = \frac{\rho}{3}. \label{apa-3}
\end{equation}
From the entropy density $s = \left(\rho + p\right)/T_{LL}$ we build the entropy flux
\begin{equation}
S_0^{\mu} = su^{\mu} = p \beta^{\mu}_{LL} - \beta_{LL\nu}T^{\mu\nu}_{0}\label{apa4}
\end{equation}
with $\beta^{\mu}_{LL} = u^{\mu}/T_{LL}$ and where subindex ${LL}$ refers to Landau-Lifshitz frame. 
The differential form for the first law, $ds = d\rho/T_{LL}$, implies
\begin{equation}
dS_0^{\mu} = -\beta_{LL\nu} dT_{0}^{\mu\nu},\label{apa5}
\end{equation}
which gives that an ideal fluid flows with no entropy production, i.e.,
\begin{equation}
S^{\mu}_{0;\mu} = -\beta_{LL\nu}T_{0;\mu}^{\mu\nu}=0.\label{apa6}
\end{equation}
Besides from $p=\rho/3$ we have $s= 4\rho/3T_{LL}$ and $ds/d\rho = 1/T_{LL}$ and we get
$\rho = \sigma\,T^4_{LL}$, where $\sigma$ is a dimensionless constant which depends on the statistics and the number of species of the particles that describe the fluid. In the case of a photon gas 
$\sigma$ is the well-known Stefan-Boltzmann constant.

A real fluid departs from an ideal one in that now
\begin{equation}
T^{\mu\nu} = T^{\mu\nu}_0 + \Pi^{\mu\nu} \label{apa7}
\end{equation}
where $\Pi^{\mu\nu}$ encodes the non-ideal properties of the flow and satisfies $\Pi^{\mu\nu}u_{\nu}=0$. 
If we still consider $S^{\mu}_0$ to be the entropy flux, we now have
\begin{equation}
S^{\mu}_{0;\mu} = -\beta_{LL\nu}T^{\mu\nu}_{0;\mu} =
\beta_{LL\nu}\Pi^{\mu\nu}_{;\mu} = -\frac{u_{\nu;\mu}}{T_{LL}}
\Pi^{\mu\nu}. \label{apa8}
\end{equation}
Positive entropy production is satisfied if
\begin{equation}
\Pi_{\mu\nu} = - \eta \sigma_{\mu\nu}\label{apa9}
\end{equation}
where $\sigma^{\mu\nu}$ is the shear tensor 
\be 
\sigma^{\mu\nu}= \Delta^{\mu\sigma}\Delta^{\nu\lambda}\left( u_{\sigma;\lambda}+u_{\lambda;\sigma}
-\frac23\Delta_{\sigma\lambda}u^{\rho}_{;\rho}\right) \label{gdtt16}
\te 
and $\eta \propto T_{LL}^{3}$ is the fluid viscosity. This constitutive relation leads to Landau-Lifshitz hydrodynamics, 
namely a covariant Navier-Stokes equation, which violates causality \cite{HisLind85}. 

We may intend to solve the problem by upgrading $\Pi^{\mu\nu}$ to a dynamical variable and adopting a Maxwell-Cattaneo equation for it, 
having eq. (\ref{apa9}) as an asymptotic limit. We then write
\begin{equation}
\Pi_{\mu\nu} = - \left[\eta \sigma_{\mu\nu} 
+ \tau\dot\Pi^{\mu\nu}\right].\label{apa10}
\end{equation}
This would follow from demanding positive entropy production with an entropy production term
\begin{equation}
S^{\mu}_{;\mu} = -\frac{\Pi^{\mu\nu}}{2T_{LL}}
\left[\sigma_{\mu\nu} + \varsigma\dot\Pi_{\mu\nu}\right]\label{apa11}
\end{equation}
and identifying later on $\tau = \varsigma\eta$. There arises the problem of what is $S^{\mu}$. A natural choice would be
\begin{equation}
S^{\mu} = S_0^{\mu} - \frac{\varsigma}{4\,T_{LL}}u^{\mu}\Pi^{\gamma\nu}\Pi_{\gamma\nu} \label{apa12}
\end{equation}
which is thermodynamically satisfactory, but leads to
\begin{equation}
S^{\mu}_{;\mu} = -\frac{\Pi^{\mu\nu}}{2T_{LL}}
\left[\sigma_{\mu\nu} + \varsigma\dot\Pi_{\mu\nu}\right] -
\frac{\varsigma}{4}\Pi^{\gamma\nu}\Pi_{\gamma\nu}\beta^{\mu}_{LL;\mu}.
\label{apa13}
\end{equation}
The extra term may be expected to be small, as it is of third order in deviations from equilibrium, 
but it is not nonnegative definite, and so we cannot be certain that the Second  Law is properly enforced. 
To guarantee that it is, we should go to higher order in eq. (\ref{apa10}), a step that would stem from 
including a new higher order term in the expression (\ref{apa11}), and then impose a condition equivalent to 
(\ref{apa12}), and so on. In other words, in order to have a thermodynamically consistent hydrodynamics
we should enforce the Second Law order by order in deviations from equilibrium.

\subsection{Minimal conformal DTT beyond Landau-Lifshitz hydrodynamics}\label{MDTT}

Instead of patching the theory one step at a time, DTTs attempt to formulate a consistent theory in its own right by postulating 
new currents, besides $T^{\mu\nu}$, which together determine the entropy flux. In its simplest form there is only one further current, 
$A^{\mu\nu\rho}$, satisfying a divergence-type equation \cite{cal013}
\begin{equation}
A^{\mu\nu\rho}_{;\rho} = I^{\mu\nu} \label{apa14}
\end{equation}
where $I^{\mu\nu}$ is a tensor source of irreversibility. A simple count of degrees of freedom tells us that we need 5 independent 
equations to complement the 4 equations from the energy momentum conservation. We impose $A^{\mu\nu\rho}$ to be totally symmetric and 
traceless on any two indices and take the transverse, traceless part of eq. (\ref{apa14}) as providing the required equations. 

The big assumption of DTTs is that we have a local First Law of the form
\begin{equation}
dS^{\mu} = -\beta_{\nu}dT^{\mu\nu} 
- \zeta_{\nu\rho}dA^{\mu\nu\rho} \label{apa15}
\end{equation}
with $\zeta_{\mu\nu}$ a new tensor variable that encodes the non-ideal properties of the flow. In particular, this leads to
\begin{equation}
S^{\mu}_{;\mu} = -\zeta_{\nu\rho}I^{\mu\nu}. \label{apa16}
\end{equation}
So the Second Law is enforced as long as
\begin{equation}
\zeta_{\mu\nu}I^{\mu\nu} \leq 0  .\label{apa17}
\end{equation}
Another consequence of eq. (\ref{apa15}) is that if we consider the Massieu function density \cite{GerLind90,GerLind91}
\be 
\Phi^{\mu}=S^{\mu}-\beta_{\nu}T^{\mu\nu}-\zeta_{\nu\rho}A^{\mu\nu\rho}\label{MDTT-1}
\te 
Then 
\bea 
\frac{\partial\Phi^{\mu}}{\partial\beta_{\nu}}&=&T^{\mu\nu},\label{MDTT-2}\\
\frac{\partial\Phi^{\mu}}{\partial\zeta_{\nu\rho}}&=&A^{\mu\nu\rho},\label{MDTT-3}
\tea
and further, the symmetry of $T^{\mu\nu}$ allows us to write 
\be 
\Phi^{\mu}=\frac{\partial\Phi}{\partial\beta_{\mu}}\label{MDTT-4}
\te 
Thus the theory is defined by specifying the scalar $\Phi$ and the tensor $I^{\mu\nu}$ as local functions of the vector $\beta^{\mu}$ 
and the tensor $\zeta^{\mu\nu}$, subjected to eq. (\ref{apa17}). Although a DTT may be derived from an underlying microscopic description, such as kinetic or field theory, when such is available, 
one of its appealing features is that one can go a long way into finding the right DTT from purely macroscopic arguments. In this
paper we shall make the simplifying assumption that the scalar $\Phi$ has no space-time dependence other than through the hydrodynamical variables themselves, for more general situations see \cite{lucas19}.

We start by writing the generating function $\Phi$ in terms of $0$th, $1$st and $2$nd order in deviations from equilibrium 

\be 
\Phi=\Phi_0+\Phi_1+\Phi_2\label{MDTT-5}
\te 
In equilibrium $\zeta^{\mu\nu}=0$, so

\be
\Phi_0=\phi_0\left( X\right) \label{MDTT-6}
\te 
where $X=\beta_{\mu}\beta^{\mu}$. Thus 

\be
\Phi^{\mu}_0=2\beta^{\mu}\phi'_0\left( X\right)\label{MDTT-7}
\te  

\be 
T_0^{\mu\nu}=4\beta^{\mu}\beta^{\nu}\phi''_0\left( X\right)+2g^{\mu\nu}\phi'_0\left( X\right)\label{MDTT-8}
\te 
then

\be 
0=T^{\mu}_{\mu}=4\left[ X\phi''_0+2\phi'_0\right] \label{MDTT-9}
\te 
Thus the only choice is 
\bea 
\phi_0&=&-\frac{\sigma }6 X^{-1}\label{MDTT-10}\\
\Phi_0^{\mu}&=&\frac{\sigma }3\beta^{\mu} X^{-2}\label{MDTT-11}\\
T_0^{\mu\nu}&=&\sigma  X^{-2}\left[ -X^ {-1}\beta^{\mu}\beta^{\nu}+\frac13\Delta^{\mu\nu}\right] \label{MDTT-12}
\tea
For the first order terms we have 
\be
\Phi_1=\phi_1\left( X\right)\zeta^{\lambda}_{\lambda}+\phi_2\left( X\right)\zeta_{\lambda\sigma}\beta^{\lambda}\beta^{\sigma}
\label{MDTT-13}
\te 
\be
\Phi_1^{\mu}=2\beta^{\mu}\left[ \phi'_1\left( X\right)\zeta^{\lambda}_{\lambda}+\phi'_2\left( X\right)\zeta_{\lambda\sigma}
\beta^{\lambda}\beta^{\sigma}\right] +2\phi_2\left( X\right)\zeta^{\mu}_{\sigma}\beta^{\sigma}\label{MDTT-14}
\te 
whereby
\bea
T^{\mu\nu}_1&=&4\beta^{\mu}\beta^{\nu}\left[ \phi''_1\left( X\right)\zeta^{\lambda}_{\lambda}+\phi''_2\left( X\right)
\zeta_{\lambda\sigma}\beta^{\lambda}\beta^{\sigma}\right] +2g^{\mu\nu}\left[ \phi'_1\left( X\right)\zeta^{\lambda}_{\lambda}
+\phi'_2\left( X\right)\zeta_{\lambda\sigma}\beta^{\lambda}\beta^{\sigma}\right]\nn
&+&4\phi'_2\left( X\right)\left( \zeta^{\mu}_{\sigma}\beta^{\nu}+\zeta^{\nu}_{\sigma}\beta^{\mu}\right) \beta^{\sigma}
+2\phi_2\left( X\right)\zeta^{\mu\nu}\label{MDTT-15}
\tea
and 
\be
A^{\mu\nu\rho}_1=2\beta^{\mu}\left[ \phi'_1\left( X\right)g^{\nu\rho}+\phi'_2\left( X\right)\beta^{\nu}\beta^{\rho}\right]
+\phi_2\left( X\right)\left( g^{\mu\nu}\beta^{\rho}+ g^{\mu\rho}\beta^{\nu}\right) \label{MDTT-16}
\te 
To obtain the right number of degrees of freedom for a conformal fluid we ask that for physically meaningful fields 
$\zeta^{\lambda}_{\lambda}= \zeta_{\lambda\sigma}\beta^{\sigma} =0$. When these conditions hold we shall say we are ``on shell''. 
As $A^{\mu\nu\rho}$ must be symmetric on any pair of indices we have that $\phi_2=2\phi'_1$,
and by demanding $A^{\mu\nu\rho}$ to be traceless on any pair of indices $X\phi'_2+3\phi_2=0$. So

\be
\phi_2=-\frac{a}{2}X^{-3} \label{MDTT-17}
\te 
with $a$ a constant that sets the intensity of the first order deviations around equilibrium, whose precise value will depend on 
the specific system under study, and 

\bea
\Phi_1^{\mu}&=&0\label{MDTT-18}\\
T^{\mu\nu}_1&=&-aX^{-3}\zeta^{\mu\nu}\label{MDTT-19}\\
A^{\mu\nu\rho}_1&=&3aX^{-4}\beta^{\mu}\beta^{\nu}\beta^{\rho}-\frac12aX^{-3}\left( g^{\mu\nu}\beta^{\rho}
+ g^{\mu\rho}\beta^{\nu}+g^{\nu\rho}\beta^{\mu}\right) \label{MDTT-20}
\tea
In writing second order terms, we leave out terms that do not contribute to the currents ``on shell''. This leaves 
\be
\Phi_2=\phi_3\left( X\right)\zeta^{2\lambda}_{\lambda}+\phi_4\left( X\right)\zeta^2_{\lambda\sigma}\beta^{\lambda}\beta^{\sigma}
\label{MDTT-21}
\te
and 
\begin{equation}
 \Phi_2^{\mu} = 2\phi_3^{\prime}(X)\zeta^{2\lambda}_{\lambda}\beta^{\mu} + 
 2\phi_4^{\prime}(X) \zeta^2_{\lambda\sigma}\beta^{\lambda}\beta^{\sigma}\beta^{\mu}
 + \phi_4(X)\zeta^2_{\lambda\sigma}\left[g^{\lambda\mu}\beta^{\sigma} + g^{\sigma\mu}\beta^{\lambda}\right]
 \label{phi-2}
\end{equation}
Therefore
\begin{eqnarray}
 T^{\mu\nu}_2 &=& 4\phi_3^{\prime\prime}(X)\zeta^{2\lambda}_{\lambda}\beta^{\mu}\beta^{\nu}
 + 2\phi^{\prime}_3(X)\zeta^{2\lambda}_{\lambda} g^{\mu\nu}
 + \phi_4(X) \zeta^2_{\lambda\sigma}\left[g^{\lambda\mu}g^{\sigma\nu} + g^{\lambda\nu}g^{\sigma\mu}\right] \label{EMT2}
\end{eqnarray}
and 
\begin{equation}
 A_2^{\mu\rho\sigma} = 8\phi_3^{\prime}(X) \zeta^{\rho\sigma}\beta^{\mu}
 + 2\phi_4(X) \left[\zeta^{\rho\mu}\beta^{\sigma} + \zeta^{\sigma\mu}\beta^{\rho}\right]
 \label{MDTT-22}
\end{equation}
From $T^{\mu}_{2\mu}=0$ we get 
\be 
4X\phi''_3+8\phi'_3+2\phi_4=0 \label{MDTT-23}
\te
From symmetry of $A^{\mu\nu\rho}$ we get $\phi_4=4\phi'_3$. So we can write
\bea
\Phi_2^{\mu}&=&\frac{b}{4}X^{-4}\beta^{\mu}\zeta^{2\lambda}_{\lambda}\nn
T^{\mu\nu}_2&=&bX^{-4}\left[ \zeta^{2\mu\nu}+\frac14\left(g^{\mu\nu}-8X^{-1}\beta^{\mu}\beta^{\nu}\right)
\zeta^{2\lambda}_{\lambda}\right] \label{MDTT-24}\\
A^{\mu\nu\rho}_2&=&\frac{b}{2}X^{-4}\left( \zeta^{\mu\nu}\beta^{\rho}+ \zeta^{\mu\rho}\beta^{\nu}+\zeta^{\nu\rho}\beta^{\mu}\right) 
\label{MDTT-25}
\tea
with $b$ a constant that sets the amplitude of the second order deviations around equilibrium. Its precise value
will depend on the specific system under study. The energy density then is
\be 
\rho=-X^{-1}\beta_{\nu}\beta_{\mu}T^{\mu\nu}= X^{-2}\left[\sigma  +\frac74bX^{-2}\zeta^{2\lambda}_{\lambda}\right]\label{MDTT-26}
\te
and the entropy current reads
\be 
S^{\mu}=\frac{4}3\sigma \beta^{\mu} X^{-2}+\frac32bX^{-4}\beta^{\mu}\zeta^{2\lambda}_{\lambda}\approx \frac43\rho\beta^{\mu}
\left[ 1-\frac58\frac b{\sigma}X^{-2}\zeta^{2\lambda}_{\lambda}\right] \label{MDTT-27}
\te
We now define
\bea 
\zeta^{\mu\nu}&=&-\sqrt{\frac{\sigma }b}XZ^{\mu\nu}\label{MDTT-28}\\
a&=&\alpha\sqrt{b\sigma } \label{MDTT-29}
\tea
where $\alpha$ is to be defined below, to get 
\bea 
T^{\mu\nu}&=&\sigma  X^{-2}\left[ -X^ {-1}\beta^{\mu}\beta^{\nu}+\frac13\Delta^{\mu\nu}+\alpha Z^{\mu\nu}
+ Z^{2\mu\nu}\right.\nn
&+&\left.\frac14\left(g^{\mu\nu}-8X^{-1}\beta^{\mu}\beta^{\nu}\right)Z^{2\lambda}_{\lambda}\right] \label{MDTT-30} \\
A^{\mu\nu\rho}&=&\frac12aX^{-4}\left\{ 6\beta^{\mu}\beta^{\nu}\beta^{\rho}
 -X\left[g^{\mu\nu}\beta^{\rho}+ g^{\mu\rho}\beta^{\nu}+g^{\nu\rho}\beta^{\mu}\right.\right.\nn
 &+& \left.\left. \alpha^{-1} 
 \left(Z^{\mu\nu}\beta^{\rho}+ Z^{\mu\rho}\beta^{\nu}+Z^{\nu\rho}\beta^{\mu}\right)\right] \right\} \label{MDTT-31}\\
S^{\mu}&=&\frac{4}3\sigma \beta^{\mu} X^{-2}\left[ 1+\frac98Z^{2\lambda}_{\lambda} \right] \label{MDTT-32}
\tea
These constitutive relations define the theory. By comparing to the Landau-Lifshitz theory above we see that on dimensional grounds 
we may write $X^{-1}=-T^2$, where $T$ has dimensions of temperature, while  $Z^{\mu\nu}$ is dimensionless. Writing $\beta^{\mu}=u^{\mu}/T$ 
the constitutive relations take the form
\bea 
T^{\mu\nu}&=&\sigma  T^4\left[ \left(1+\frac74Z^{2\lambda}_{\;\;\lambda}\right)\left(u^{\mu}u^{\nu}+\frac13\Delta^{\mu\nu}\right)
+\alpha Z^{\mu\nu}+ \left(Z^{2\mu\nu}-\frac13\Delta^{\mu\nu}Z^{2\lambda}_{\;\;\lambda}\right)\right] \label{current-emt}\\
A^{\mu\nu\rho}&=&\frac12aT^{5}\left\{ 6u^{\mu}u^{\nu}u^{\rho}+\left[ g^{\mu\nu}u^{\rho}+ g^{\mu\rho}u^{\nu}+g^{\nu\rho}u^{\mu}
\right.\right.\nn
&+&\left.\left.\alpha^{-1} \left(Z^{\mu\nu}u^{\rho} + Z^{\mu\rho}u^{\nu}  + Z^{\nu\rho}u^{\mu}\right)\right] \right\} 
 \label{current-a}\\
S^{\mu}&=&\frac{4}3\sigma u^{\mu} T^3\left[ 1+\frac98Z^{2\lambda}_{\lambda} \right] 
\label{current-s}
\tea

To fix the remaining constants we ask that the theory reproduces Landau-Lifshitz hydrodynamics to first order in deviations 
from equilibrium. This requires $T=T_{LL}\left(1+O\left( Z^2\right) \right) $ and 
\be 
\sigma \alpha T^4Z^{\mu\nu}=-\eta \sigma^{\mu\nu}
\label{fopi}
\te
In the DTT framework, eq. (\ref{fopi}) ought to follow from the transverse, traceless part of the first-order conservation law for 
$A^{\mu\nu\rho}$ 
\be 
I^{\mu\nu}=\Lambda^{\mu\nu}_{\sigma\lambda} A^{\sigma\lambda\rho}_{;\rho}= \frac12 a T^5\sigma^{\mu\nu} \label{MDTT-33}
\te
where 
\be 
\Lambda^{\mu\nu}_{\sigma\lambda}=\frac12\left[\Delta^{\mu}_{\sigma}\Delta^{\nu}_{\lambda}+ \Delta^{\mu}_{\lambda}\Delta^{\nu}_{\sigma}
-\frac23\Delta^{\mu\nu}\Delta_{\sigma\lambda}\right]
\label{tenproy}
\te
is the complete transverse and traceless spatial projector. Therefore we must have 
\begin{equation}
I^{\mu\nu} = - \frac{a\sigma \alpha T^6}{2}\frac{T^3}{\eta}Z^{\mu\nu}\label{apb15}
\end{equation}
It is convenient to introduce the relaxation time $\tau$ through the Anderson-Witting prescription \cite{AndWit,AndWit2}
\be 
I^{\mu\nu}=-\frac{aT^5}{2\alpha\tau}Z^{\mu\nu} \label{MDTT-34}
\te 
whereby 
\be 
\alpha^2=\frac1{\sigma T\tau}\left( \frac{\eta}{T^3}\right) 
\label{alfasq}
\te 
We may estimate  $\eta$ from the AdS/CFT bound \cite{PolSonStar01}, $\eta\ge \left(4/3\right)\sigma T_0^3/4\pi$,  
$T_0$ being a fiducial equilibrium temperature. In the next subsection we show that causality requires $\alpha^2\le 2/3$ 
and so $T\tau\ge 3\eta/2\sigma T_0^3 \ge 1/2\pi$. In what follows we shall be interested in the ``free-streaming'' 
limit $\tau\to\infty$, thus $\alpha\to 0$.

As we shall show in next section, when thermal fluctuations are considered $I^{\mu\nu}$ acquires a stochastic component, 
$I^{\mu\nu}\to I^{\mu\nu}+F^{\mu\nu}$.  $F^{\mu\nu}$ is a stochastic source which may be derived from fluctuation-dissipation 
considerations and will be described in more detail below. Observe that this force sources entropy and not energy, as there is 
no stirring in the equations for the conservation of $T^{\mu\nu}$. 

\subsection{Propagation speeds and ``incompressibility''}\label{incom}
Before we proceed, we shall derive the propagation speeds for different types of linearized fluctuations around equilibrium \cite{M99}, 
and show that in the $\alpha\to 0$ limit scalar modes may be regarded as frozen, thus the fluid behaves as an incompressible one. 

We are looking at a situation where a discontinuity is propagating along the surface $z=ct$, $c$ being the desired propagation speed. 
Above the front the fluid is in equilibrium, so $T=T_0=$ constant, $u^{\mu}=U^{\mu}=\left( 1,0,0,0\right) $ and $Z^{\mu\nu}=0$. 
Hydrodynamic variables are continuous across the front. Observe that any variable $\mathcal{X}$ which remains constant at the front 
must obey $\dot {\mathcal{X}}=-c\mathcal{X}'$, where $\mathcal{X}'=\mathcal{X}_{,3}$, where subindex $3$ refers to the $z$ coordinate. 
Moreover the conditions $u^2=-1$ and  $Z^{\mu\nu}u_{\nu}=0$ show that $u^0_{,\rho}=Z^{\mu 0}_{,\rho}=0$, and the condition 
$Z^{\rho}_{\rho}=0$ shows that $Z^a_a=-Z^3_3$ (indices $a,b$ run from $1$ to $2$, which denote $x$ and $y$ coordinates). 

The theory decomposes into tensor modes $Z^{ab}+\left( 1/2\right) \delta^{ab}Z^3_3$, which do not propagate, vector modes $u^a$ and $Z^{a3}$ and scalar modes $T$, $u^3$ and $Z^{33}$. Writing the 
conservation equations  $T^{\mu\nu}_{,\nu}=0$ and 
$\Lambda^{\mu\nu}_{\rho\lambda}A^{\rho\lambda\sigma}_{,\sigma}=0$ (since $I^{\mu\nu}=0$ at the front), and eliminating time derivatives, we get two sets of equations. For the vector modes 
\bea 
u'_a-\frac c{\alpha}Z'_{a3}&=&0\label{incom-1} \\
-\frac43 cu'_a+\alpha Z'_{a3}&=&0 \label{incom-2}
\tea
which shows that vector modes propagate with speed 
\be
c_V^2=3\alpha^2/4 \label{incom-3}
\te 
For scalar modes we get 
\bea 
\frac43 cu'_3-\frac c{\alpha}Z'_{33}&=&0 \label{incom-4}\\
\frac{T'}{T_0}-cu'_3+\frac34\alpha Z'_{33}&=&0 \label{incom-5}\\
-c\frac{T'}{T_0}+\frac13 u'_3&=&0 \label{incom-6}
\tea 
which admits a non-propagating mode with $u'_3=0$, $T'/T_0=-\left( 3/4\right)\alpha Z'_{33}$, and two propagating modes with speed
\be 
c_S^2=\frac13+\alpha^2 \label{incom-7}
\te 
So causality demands $\alpha^2\le 2/3$, and $c_S\gg c_V$ when $\alpha^2 \ll 2/3$. This means that the only interaction of interest
is between tensor and vector modes.

\section{The Martin-Siggia-Rose effective action}\label{MSREA}

As stated above, we shall be interested in fluids stirred by their own thermal fluctuations. Therefore in this section we shall review
the fluctuation dissipation theorem appropriate to causal relativistic real fluids \cite{cal98} (see also \cite{JouRubiCV80,murase19}), and then use the MSR formulation \cite{msr-73,dedo-76,kam-11,eyink96,zancal02,mcomb14} to develop an effective action from where we 
can get the Dyson equations for the stochastic correlations of vector and tensor hydrodynamic variables.

\subsection{Fluctuation-dissipation theorem in a DTT framework}\label{fdtdtt}

In this subsection we summarize the derivation of the fluctuation-dissipation theorem as applied to causal relativistic fluid theories.
Following Ref. \cite{cal98} we define the following shorthand notation for the variables described above
\begin{eqnarray}
X^A&=&(\beta^a,\zeta^{ab}),\label{fdtdtt-1} \\
A^a_A&=&(T^{ab},A^{abc}),\label{fdtdtt-2}\\
I_B&=&(0,I_{ab})\,\label{fdtdtt-3}\\
F_B&=&(0,F_{ab}),\label{fdtdtt-4}\\
S^a&=&\Phi^a-X^BA^a_B, \label{fdtdtt-5}\\
A^a_{B}&=&\frac{\delta \Phi^a}{\delta X^B},\label{fdtdtt-6}
\end{eqnarray}
with $\Phi^a$ the vector generating functional. The entropy production is given by
\be
S^a_{;a}=-X^BA^a_{B;a}. \label{fdtdtt-7}
\te
To include thermal fluctuations we add random sources $F_B$ in the equations of motion which then become Langevin-type equations, namely
\be
A^a_{B;a}=I_B+F_B\label{equationsofmotion}.
\te
A satisfactory theory must predict vanishing mean entropy production in equilibrium, so
\be
\langle S^a_{;a}\rangle= -\langle X^B(x)I_B(x)\rangle - \langle X^B(x)F_B(x)\rangle=0. \label{fdtdtt-8}
\te
However, because the coincidence limit may not be well defined, we impose a stronger condition due to elementary causality considerations, which is
\be
\langle X^B(x)I_B(x')\rangle+\langle X^B(x)F_B(x')\rangle=0
\label{nullmeanentropyproduction}
\te
for every space-like pair $(x,x')$. In the following we shall assume that we have defined the time in such a way that $x$ and $x'$ belong to 
the same equal time surface, namely $t=t'$. In the linear approximation $I_B$ is a linear function of $ X^C$, then 
\be
\langle X^B(x)I_B(x')\rangle = \int d^3x'' I_{(B,C)}\langle X^B(x) X^C(x'')\rangle.\label{zetai1}
\te
with
\begin{equation}
 I_{(B,C)} = \frac12 \left[\frac{\delta I_B(x')}{\delta  X^C(x'')}+\frac{\delta I_C(x'')}{\delta  X^B(x')}\right].
 \label{IBC-def}
\end{equation}
Only the symmetrized derivative occurs in (\ref{zetai1}) due to the symmetry of the stochastic average. Assuming Gaussian white noise
\be
\langle F_B(x)F_C(x')\rangle=\sigma_{BC}(x,x')\,\delta(t-t'), \label{fdtdtt-9}
\te
we have that the correlations between fluxes and noise is
\be
\langle X^A(x)F_B(x')\rangle=\int d^3x''\sigma_{BC}(x',x'')\;\frac{\delta X^A(x)}{\delta F_C(x'')}.\label{zetaj}
\te
where $t''=t'$. The fluctuation dissipation theorem follows from 
\be
\langle X^B(x) X^C(x')\rangle=2\;\frac{\delta X^B(x)}{\delta F_C(x')}=2\;\frac{\delta X^C(x')}{\delta F_B(x)},
\label{zzzj}
\te
whenever $t=t'$, which implies
\be
\langle X^B(x)I_B(x')\rangle =\int d^3x'' I_{(B,C)}\, \frac{\delta X^B(x)}{\delta F_C(x'')}.
\label{zetai}
\te

Using (\ref{nullmeanentropyproduction}), (\ref{IBC-def}), (\ref{zetaj}) and (\ref{zetai}) we get
\be
\sigma_{BC}=-2\;I_{(B,C)}, \label{fdtdtt-10}
\te
or equivalently
\be
\langle F_A(x)F_B(x')\rangle=-\left[\frac{\delta I_A(x)}{\delta  X^B(x')}+\frac{\delta I_B(x')}{\delta  X^A(x)}\right]\delta (t-t').
\label{fdtdtt-11}
\te
We use this version of the fluctuation-dissipation theorem in order to set the correlation function of the noise source. 
Of course, as we show in the main text, in the limit in which our DTT converges to the Landau-Lifshitz hydrodynamics, 
the correlation of the stochastic energy-momentum tensor converges to the well-known Landau-Lifshitz noise \cite{LLflucthyd,LLsmII}.

To verify Eq. (\ref{zzzj}), let us multiply both sides by the non-singular matrix
\be
M_{AB}=n_a\frac{\delta^2\,\Phi^a}{\delta X^A\delta X^B}\label{fdtdtt-12}
\te
where $n_a$ is the unit normal field to the equal time surface containing both $x$ and $x'$. In the linear approximation $\Phi^a$ is 
quadratic on $ X^A$ and 
\be
M_{AB} X^B=\frac{\delta (n_a\Phi^a)}{\delta X^A}=-\frac{\delta \Phi^0}{\delta X^A}. \label{fdtdtt-13}
\te
In equilibrium we may apply the Einstein's formula, relating the thermodynamic potentials to the distribution function of fluctuations, to conclude that
\be
M_{AB}\langle X^B(x) X^C(x')\rangle=-\delta^C_A\,\delta^{(3)}(x,x')\label{leftside}
\te
where $\delta^{(3)}(x,x')$ is the three-dimensional covariant delta function on the Cauchy surface. This is a generalized version of the 
equipartition theorem. On the other hand
\be
M_{AB}\frac{\delta X^B(x)}{\delta F_C(x')}=-\frac{\delta A^0_A(x)}{\delta F_C(x')}
\label{rightside}
\te
with $A^0_A=-n_aA^a_A$. It is possible to write the equations of motion (\ref{equationsofmotion}) as
\be
\frac{\partial A^0_A(x)}{\partial t}+L_A(x)=F_A(x) \label{fdtdtt-14}
\te
where $L_A$ involves the field variables on the surface, but not their normal derivatives, and 
$\partial/\partial t:=n^a\;\partial/\partial x^a$. Indeed
\be
\frac{\delta A^0_A(x)}{\delta F_C(x')}=\frac12\delta^C_A\delta^{(3)}(x,x'). \label{fdtdtt-15}
\te
The factor $1/2$ takes into account the average of the derivative evaluated in $x=x'^-$ and $x=x'^+$. Therefore (\ref{leftside}) 
and (\ref{rightside}) are equal. Due to the non-singularity of $M_{AB}$, equation (\ref{zzzj}) holds.

In the case at hand, these results imply that only the equation for $A^{\mu\nu\rho}$ acquires a random source, and then
\be
\left\langle F^{\mu\nu}\left(x\right)F_{\sigma\lambda}\left(x'\right)\right\rangle 
= N\delta(t-t')\delta^{\left( 3\right) }(x-x')\Lambda^{\mu\nu}_{\sigma\lambda} \label{fdtdtt-16}
\te 
where

\be
N=\frac{a^2T_0^7}{\alpha^2\sigma \tau} \label{ff-ns}
\te

\subsection{MSR and the 2PI Effective action}\label{MSR2PIEA}

The main tools to study the evolution and physical
properties of a stochastic system are their different propagators,
or Green functions, because they determine the
response of the system to its own thermal fluctuations as well 
as to the fluctuations in the initial conditions
\cite{KMPST19a,KMPST19b,KMPOS20}. In the regime of strong 
fluctuations this study involves corrections due to non-linear
effects \cite{wyld61,lee65,mcomb90,mcomb14} and it is in this
scenario where effective field theory methods such as the
MSR show their power \cite{msr-73,dedo-76,kam-11,zancal02};
for applications to theories of turbulence see
\cite{wyld61,lee65,FoNeSte77}. It was also used recently 
to study fluctuations in relativistic
Landau-Lifshitz theory  \cite{Kovtun12,KMR14}.

We now proceed with the analysis of the correlations in the 
theory. From the analysis of the propagation velocities we know that in the free-streaming regime the scalar modes propagate faster 
than the vector ones and therefore can be considered as frozen. 
In other words, in the considered limit the flow may be regarded as
``incompressible'' (cfr. subsection (\ref{incom})). The MSR formalism will allow to convert the problem of classical fluctuations into a quantum field theory one, for which we shall derive the 
Two-Particle-Irreducible
Effective Action (2PIEA). This formalism yields the Schwinger-Dyson equations for the propagators in the most direct way.

Before going on an important remark is in order. In hydrodynamics there is no explicit single `small' parameter, such as $\hbar$ in 
quantum field theory, which can be used to organize the perturbative expansion. For this reason it has been propossed that the loop 
expansion should be understood as an expansion in `the complexity of the interaction' \cite{wyld61,lee65}, since due to the randomness of the stirring, the sum of higher order terms will tend to cancel. 
On the other hand, it is possible to identify the relevant small parameters in the theory through the scaling behavior of restricted sets of graphs. In the case at hand, this analysis suggests that the loop expansion is an expansion in powers of 
$\left(p/p_{\mathrm{L}}\right)^3$ where
\be
p_{\mathrm{L}}=\left(c_V^2\,\sigma \,T_0^3\right)^{1/3}.\label{ploop}
\te
(see Appendix \ref{app:loopscaling}). In consequence the loop
expansion is consistent while $p<p_{\mathrm{L}}$.

Let us return to the construction of the 2PIEA. We continue to use the abridged notation from eqs. (\ref{fdtdtt-1})-(\ref{fdtdtt-6}). The equations of motion have the form $P_A=F_A$, 
where the $P$'s are the left hand sides of the EMT $T^{\mu\nu}$ and the nonequilibrium current $A^{\mu\nu\rho}$ conservation equations. 
If the sources $F_A$ are given, we call $X^{ A}\left[F\right]$ the solution to the equations. Under thermal noise, all 
$\left\langle X^{ A}\right\rangle=0$. Therefore we can write a generating functional for the correlation functions 
$\left\langle X^{ A}X^{\beta}\right\rangle$
\be
e^{iW\left[K_{ A B}\right]}=\int \;DX^{ A}\;\int \;DF_A\;\mathcal{P}\left[F^A\right]e^{i\int\;K_{A B}X^{ A}X^{ B}/2}
\delta\left(X^A-X^A\left[F\right]\right) \label{MSR2PIEA-1}
\te
where $\mathcal{P}\left[F^A\right]$ is the Gaussian probability density for the sources, $K_{AB}$ are the currents introduced 
in the formalism to couple to the variables of the theory and the integration is performed over all the noise realizations. Observe that
\be
\delta\left(X^A-X^A\left[F\right]\right)=\mathrm{Det}\left[\frac{\delta P_A}{\delta X^{ A}}\right]\delta\left(P_A-F_A\right)
\label{MSR2PIEA-2}
\te
where the determinant can be proved to be a constant \cite{zinnjustin} and will be consequently disregarded. We exponentiate the delta function by adding auxiliary fields $Y_A$
\be
e^{iW\left[K_{ A B}\right]}=\int \;DY^A\;\int \;DX^{ A}\;\int \;DF_A\;\mathcal{P}\left[F^A\right]e^{i\int\;K_{ A B}X^{ A}X^{ B}/2}
e^{i\int\;Y^A\left(P_A-F_A\right)} \label{MSR2PIEA-3}
\te
Introducing the source correlations
\be
\left\langle F_AF_B\right\rangle=N_{AB} \label{MSR2PIEA-4}
\te
we finally obtain
\be
e^{iW\left[K_{ A B}\right]}=\int \;DY^A\;\int \;DX^{ A}\;e^{iS}e^{i\int\;K_{ A B}X^{ A}X^{ B}/2} \label{MSR2PIEA-5}
\te
where
\be
S=\int\; d^4x\left[Y^AP_A+\frac i2Y^AN_{AB}Y^B\right]
\label{class}
\te
In fact, we have mapped the stochastic hydrodynamic problem into a nonequilibrium field theory one, where $S$ from eq. (\ref{class}) 
plays the role of ``classical'' action. We may formally add new sources coupled to the auxiliary fields and consider the whole string 
$\mathcal{X}^K=\left(X^{ A},Y^A\right)$ as degrees of freedom of the theory.

The Legendre transform of the generating function is the 2PIEA 
$\Gamma\left[\mathcal{G}^{JK}\right]$, where the 
$\mathcal{G}^{JK}=\left\langle \mathcal{X}^J\mathcal{X}^K\right\rangle$ are the thermal correlations we seek. Once the 2PIEA is known, the actual correlations are obtained as extrema
\be
\frac{\delta\Gamma}{\delta\mathcal{G}^{JK}}=0
\label{SD}
\te
The 2PIEA has the structure \cite{calhu08}
\be
\Gamma =\frac12\left.\frac{\delta^2S}{\delta\mathcal{X}^{J}\delta\mathcal{X}^{J}}\right|_{\mathcal{X}=0}\mathcal{G}^{JK}
-\frac i2\ln\;\mathrm{Det}\left[\mathcal{G}^{JK}\right]+\Gamma_{2Q}
\label{2PIEA}
\te
where $\Gamma_{2Q}$ is the sum of all two-particle irreducible Feynman graphs for a theory whose interactions are the terms cubic or 
higher in $S$, and carrying propagators $\mathcal{G}^{JK}$. 

It so happens that $\left\langle Y^AY^B\right\rangle =0$, and also
\be
\left.\frac{\delta^2S}{\delta {X}^{ A}\delta {X}^{ B}}\right|_{\mathcal{X}=0}=
\frac{\delta\Gamma_{2Q}}{\delta\left\langle {X}^{ A} {X}^{ B}\right\rangle}=0
\label{MSR2PIEA-6}
\te
so actually we get two sets of equations, one for the retarded propagators $G_{ret}^{ A A}=-i\left\langle X^{ A}Y^A\right\rangle$
\be
\left\{\left.\frac{\delta^2S}{\delta Y^B\delta X^{ A}}\right|_{\mathcal{X}=0}
+2\frac{\delta\Gamma_{2Q}}{\delta \left\langle Y^B {X}^{ A}\right\rangle}\right\}G_{ret}^{AA}=\delta^A_B
\label{retarded}
\te 
and another for the actual thermal correlations $G^{AB}=\left\langle X^{ A}X^{ B}\right\rangle$
\be
\left\{\left.\frac{\delta^2S}{\delta Y^B\delta X^{ A}}\right|_{\mathcal{X}=0}+2\frac{\delta\Gamma_{2Q}}{\delta \left\langle Y^B {X}^{ A}
\right\rangle}\right\}G^{AB}+
i\left\{\left.\frac{\delta^2S}{\delta Y^B\delta Y^C}\right|_{\mathcal{X}=0}
+2\frac{\delta\Gamma_{2Q}}{\delta\left\langle Y^B Y^C\right\rangle}\right\}G_{adv}^{CB}=0
\label{Hadamards}
\te 
where $G_{adv}^{C B}=-i\left\langle Y^CX^{ B}\right\rangle$, with the integral
\be
G^{A B}=\left( -i\right) G_{ret}^{ A A}G_{adv}^{B B }\left\{\left.\frac{\delta^2S}{\delta Y^A\delta Y^B}\right|_{\mathcal{X}=0}
+2\frac{\delta\Gamma_{2Q}}{\delta \left\langle Y^A Y^B\right\rangle}\right\}
\label{correlations}
\te
\subsection{Induced dynamics}\label{idtm}
We now begin to investigate the induced dynamic in the presence of thermal fluctuations. As we have seen, it is given by 
eqs. (\ref{retarded}) for the causal, or retarded correlators and (\ref{Hadamards}) for the symmetric, or Hadamard two-point functions. 
In equilibrium we have $\left\langle \mathcal{X}^K\right\rangle=0$. The MSR ``classical'' action  (\ref{class}) may be written as
\be
S_C\left[\mathcal{X}^K\right]=S_Q\left[\mathcal{X}^K\right]+S_I\left[\mathcal{X}^K\right] \label{idtm-1}
\te
where $S_Q$ is quadratic and $S_I$ contains the interaction terms; 
in our case  we only keep terms cubic in the fields in $S_C$. 
The 2PIEA is given by eq. (\ref{2PIEA}), with
\be
e^{i\Gamma_{2Q}}=\mathcal{N}\int\;D\mathcal{X}^K\;e^{-\frac12\mathcal{X}^K\left(G^{-1}\right)_{KL}\mathcal{X}^L
+iS_C\left[\mathcal{X}^K\right]} \label{idtm-2}
\te 
and where the $G^{KL}$ are the propagators
\be
G^{KL}=\left\langle \mathcal{X}^K\mathcal{X}^L\right\rangle \label{idtm-3}
\te
and $\mathcal{N}\propto\left(\mathrm{Det}G^{KL}\right)^{-1/2}$ \cite{calhu08}.
We use the notation
\be
\left\langle \mathcal{O}\right\rangle=\mathcal{N}\int\;D\mathcal{X}^K\;e^{-\frac12\mathcal{X}^K\left(G^{-1}\right)_{KL}\mathcal{X}^L}
\mathcal{O}\left[\mathcal{X}^K\right] \label{idtm-4}
\te
The normalization is set up so that $\left\langle 1\right\rangle=1$. If, as in our case, 
$\left\langle S_C\right\rangle=0$, then the lowest order contribution to $\Gamma_{2Q}$ is
\be
\Gamma_{2Q}=\frac i2\left\langle S_C^2\right\rangle \label{idtm-5}
\te
and then the self energies read
\be
\Sigma_{KL}=2\frac{\delta\Gamma_{2Q}}{\delta G^{KL}}=i\left\langle \frac{\delta S_C}{\delta \mathcal{X}^K}
\frac{\delta S_C}{\delta \mathcal{X}^L}\right\rangle \label{idtm-6}
\te
The expectation value on the r.h.s. is developed in terms of Feynman graphs with propagators $G^{KL}$ in the internal legs, and where 
only 2PI graphs are considered \cite{calhu08}. Again to lowest order, 
we may replace the full propagators by their lowest order approximations
\be
\frac{\delta^2S_Q}{\delta\mathcal{X}^K\delta\mathcal{X}^M}G_0^{ML}=i\delta^L_K \label{idtm-7}
\te
which describe the correlations of linearized fluctuations around equilibrium. The equations for nonlinear fluctuations
(\ref{retarded}) and (\ref{Hadamards}) can then be written in compact form as
\be
\left[\frac{\delta^2S_Q}{\delta\mathcal{X}^K\delta\mathcal{X}^M}+\Sigma_{KM}\right]G^{ML}=i\delta^L_K \label{idtm-8}
\te

\section{Linear fluctuations around equilibrium}\label{LFAE}
From the discussion above, to formulate a MSR effective action for the minimal DTT, we consider a ``classical'' action of the form 
(cfr. eq. (\ref{class}))
\be
S=\frac i2N\int\;d^4x\;Y_{\mu\nu}Y^{\mu\nu}-\int\;d^4x\;\left\{Y_{\mu;\nu}T^{\mu\nu}+Y_{\mu\nu;\rho}A^{\mu\nu\rho}
+Y_{\mu\nu}I^{\mu\nu}\right\}
\label{class2}
\te
Since we are disregarding scalar modes, the auxiliary fields $Y_{\mu}$ and $Y_{\mu\nu}$ contain only vector and tensor degrees of freedom. 
This means that the independent variables are a three vector $Y^j$ and a three tensor $Y^{jk}$ obeying 
$Y^j_{;j}=Y^{jk}_{;j}=Y^{k}_{k}=0 $. The time components are then constrained through $Y_{\mu}u^{\mu}=Y_{\mu\nu}u^{\mu}=Y^{\mu}_{\mu}=0$. 
Observe that $Y_{\mu}$ has units of $T^{-1}$, while $Y_{\mu\nu}$ has units of $T^{-2}$. Explicitly expr. (\ref{class2}) can be
decomposed as $S=S_N+S_T+S_A+S_I$ with
\bea
S_N&=&\frac{iN}{2}\int\;d^4x\;Y_{\mu\nu}Y^{\mu\nu} \label{LFAE-1}\\
S_T&=&-\sigma T_0^4 \int\;d^4x\; Y_{\mu;\nu}\left[u^{\mu}u^{\nu}+\frac13\Delta^{\mu\nu}+\alpha Z^{\mu\nu}
+ Z^{2\mu\nu}+\frac14\left(g^{\mu\nu}-8u^{\mu}u^{\nu}\right)Z^{2\lambda}_{\lambda}\right] \label{LFAE-2}\\
S_A&=&-\frac12aT_0^5\int\;d^4x\; Y_{\mu\nu;\rho}\left[ 6u^{\mu}u^{\nu}u^{\rho}+
 g^{\mu\nu}u^{\rho}+ g^{\mu\rho}u^{\nu}+g^{\nu\rho}u^{\mu}\right.\nn
&+&\left.\frac1{\alpha} \left(Z^{\mu\nu}u^{\rho}+ Z^{\mu\rho}u^{\nu}+Z^{\nu\rho}u^{\mu}\right) \right]\label{LFAE-3}\\
S_I&=&\frac{T_0^5}{2\alpha\tau}\int\;d^4x\; Y_{\mu\nu} \;Z^{\mu\nu} \label{LFAE-4}
\tea

\subsection{Identifying the physical degrees of freedom}\label{ipdf}
We now apply the above formalism to study thermal fluctuations 
around a fiducial 
equilibrium configuration with velocity $U^{\mu} = (1,0,0,0)$ and temperature $T_0$. We shall keep only terms which are quadratic 
($S_q$) or cubic $S_c$, (not to be confused with the 'classical Action' referred above) 
in deviations from equilibrium. We shall derive the quadratic terms in the next subsection, and come back to the cubic terms 
later on, in subsection (\ref{ctca}).

We write $u^{\mu}=U^{\mu}+V^{\mu}$. The condition $u^2=-1$ becomes $U_{\mu}V^{\mu}=-V^2/2$ and so
\be
V^{\mu}=h^{\mu}_{\nu}V^{\nu}+\frac12U^{\mu}V^2 \label{ipdf-1}
\te
This suggests taking $V^k=\Delta^{k}_{\nu}V^{\nu}$ as independent variables, whereby 
\be
V^0=\frac12\left(V_kV^k-\left(V^0\right)^2\right)\approx \frac12V_kV^k+\;h.o.
\label{vinculo}
\te 
Any transverse tensor admits a similar decomposition, namely
\bea
Y_0&\approx&Y_kV^k+\;h.o.\label{ipdf-2}\\
Y_{0j}&\approx&Y_{kj}V^k+\;h.o.\label{ipdf-3}\\
Y_{00}&\approx&Y_{kj}V^kV^j+\;h.o.\label{ipdf-4}\\
Z_{0j}&\approx&Z_{kj}V^k+\;h.o.\label{ipdf-5}\\
Z_{00}&\approx&Z_{kj}V^kV^j+\;h.o. \label{ipdf-6}
\tea
where $h.o.$ means 'higher orders'. Moreover $Y^k_k=Z^k_k=0$. We can thus identify the quadratic and cubic terms in the ``classical'' 
action. The quadratic terms are
\bea
S_{Nq}&=&\frac{iN}{2}\int\;d^4x\;Y_{jk}Y^{jk}\label{ipdf-7}\\
S_{Tq}&=&-\sigma T_0^4 \int\;d^4x\;\left[ \frac43Y_{j;0}V^j +\alpha Y_{j;k}Z^{jk}\right]\label{ipdf-8}\\
S_{Aq}&=&-\frac12aT_0^5\int\;d^4x\;\left[\frac1{\alpha}  Y_{jk;0}Z^{jk}+Y_{jk;l}\left( g^{jl}V^{k}+g^{kl}V^{j}\right) \right]
\label{ipdf-9}\\
S_{Iq}&=&\frac{aT_0^5}{2\alpha\tau}\int\;d^4x\; Y_{jk} \;Z^{jk} \label{ipdf-10}
\tea
and recalling that $V^j_{;j}=Y^j_{;j}=Z^{ij}_{;ij}=Y^{ij}_{;ij}=0$, the cubic terms read 
\bea
S_{Tc}&=&-\sigma T_0^4 \int\;d^4x\; \left[-\alpha Y_{j;0}Z_{kj}V^k+Y_{j;k}\left(\frac43V^jV^k+ Z^{jl}Z^k_l\right)\right]
\label{ipdf-11}\\
S_{Ac}&=&-\frac12aT_0^5\int\;d^4x\; \left[10\left(Y_{kj}V^k\right)_{;0}V^{j}+\frac2{\alpha}\left(Y_{kj}V^k\right)_{;l}  Z^{jl} \right.\nn
&+&\left.\frac1{\alpha} Y_{jk;l} \left(Z^{jk}V^{l}+ Z^{kl}V^{j}+Z^{lj}V^k\right) \right] \label{ipdf-12}
\tea
Our next step is to Fourier transform all degrees of freedom. We adopt the convention
\be
V^k\left(\vec x, t\right)=\int\frac{d^3p}{\left(2\pi\right)^3}e^{i\vec p\vec x}V^k\left(p,t\right) \label{ipdf-13}
\te
to get
\bea
S_{Nq}&=&\frac{iN}{2}\int\;dt\frac{d^3p}{\left(2\pi\right)^3}\;Y_{jk}\left(-p,t\right)Y^{jk}\left(p,t\right)\label{ipdf-14}\\
S_{Tq}&=&-\sigma T_0^4 \int\;dt\frac{d^3p}{\left(2\pi\right)^3}\;\left[ \frac43Y_{j;0}\left(-p,t\right)V^j\left(p,t\right)
-ip_k\alpha Y_{j}\left(-p,t\right)Z^{jk}\left(p,t\right)\right]\label{ipdf-15}\\
S_{Aq}&=&-\frac12aT_0^5\int\;dt\frac{d^3p}{\left(2\pi\right)^3}\;\left[\frac1{\alpha}  Y_{jk;0}\left(-p,t\right)Z^{jk}\left(p,t\right)
-2ip^kY_{jk}\left(-p,t\right)V^{j}\left(p,t\right) \right]\label{ipdf-16}\\
S_{Iq}&=&\frac{aT_0^5}{2\alpha\tau}\int\;dt\frac{d^3p}{\left(2\pi\right)^3}\; Y_{jk} \left(-p,t\right)
\;Z^{jk}\left(p,t\right) \label{ipdf-17}
\tea
and
\bea
S_{Tc}&=&-\sigma T_0^4 \int\;dt\frac{d^3p}{\left(2\pi\right)^3}\frac{d^3q}{\left(2\pi\right)^3}\; \left[-\alpha Y_{j;0}
\left(-p-q,t\right)Z_{kj}\left(q,t\right)V^k\left(p,t\right)\right.\nn
&-&\left.i\left(p+q\right)_kY_{j}\left(-p-q,t\right)\left(\frac43V^j\left(p,t\right)V^k\left(q,t\right)+ Z^{jl}\left(p,t\right)Z^k_l
\left(q,t\right)\right)\right]\label{ipdf-18}\\
S_{Ac}&=&-\frac12aT_0^5\int\;dt\frac{d^3p}{\left(2\pi\right)^3}\frac{d^3q}{\left(2\pi\right)^3}\; \left[5Y_{kj;0}\left(-p-q,t\right)V^k
\left(p,t\right)V^{j}\left(q,t\right)\right.\nn
&-&iq_l\frac2{\alpha}Y_{kj}\left(-p-q,t\right)V^k\left(p,t\right) Z^{jl}\left(q,t\right)\nn
&-&\left.\frac i{\alpha} \left(p+q\right)_lY_{jk}\left(-p-q,t\right) \left(Z^{jk}\left(q,t\right)V^{l}\left(p,t\right)+ 2Z^{kl}
\left(q,t\right)V^{j}\left(p,t\right)\right) \right] \label{ipdf-19}\\
\tea
We may simplify these expressions by using the linear equations of motion derived from the quadratic terms, namely 
\bea 
0&=&-\sigma T_0^4 \frac43Y_{j;0}\left(-p,t\right) +aT_0^5ip^kY_{jk}\left(-p,t\right)\label{ipdf-20} \\
0&=&-\frac1{2\alpha}aT_0^5 Y_{jk;0}\left(-p,t\right)+\frac i2\alpha\sigma T_0^4 \left( p_k Y_{j}\left(-p,t\right)\right.\nn
&+&\left.p_j Y_k\left(-p,t\right)\right) +\frac{aT_0^5}{2\alpha\tau}Y_{jk} \left(-p,t\right) \label{ipdf-21}
\tea
where $O_{;0}$ refers to the time derivative, to get 
\bea
S_{Tc}&=& \int\;dt\frac{d^3p}{\left(2\pi\right)^3}\frac{d^3q}{\left(2\pi\right)^3}\; \left[\frac34\alpha aT_0^5ip^kY_{jk}
\left(-p,t\right)Z_{lj}\left(q,t\right)V^l\left(p-q,t\right)\right.\nn
&+&\left. i\sigma T_0^4p_kY_{j}\left(-p,t\right)\left(\frac43V^j\left(p-q,t\right)V^k\left(q,t\right)+ Z^{jl}\left(p-q,t\right)Z^k_l
\left(q,t\right)\right)\right]\label{ipdf-22}\\
S_{Ac}&=&-\int\;dt\frac{d^3p}{\left(2\pi\right)^3}\frac{d^3q}{\left(2\pi\right)^3}\; \left[5\left[ i\alpha\sigma T_0^4  p_k Y_{j}
\left(-p,t\right)+\frac{aT_0^5}{2\alpha\tau}Y_{jk} \left(-p,t\right) \right] V^k\left(p-q,t\right)V^{j}\left(q,t\right)\right.\nn
&-&iq_l\frac{a}{\alpha}T_0^5Y_{kj}\left(-p,t\right)V^k\left(p-q,t\right) Z^{jl}\left(q,t\right) \nn
&-&\left.\frac i{\alpha}\frac{a}2T_0^5 Y_{jk}\left(-p,t\right) p_l\left(Z^{jk}\left(q,t\right)V^{l}\left(p-q,t\right)+ 2Z^{kl}
\left(q,t\right)V^{j}\left(p-q,t\right)\right) \right] \label{ipdf-23}
\tea
Finally, we discriminate between vector and proper tensor modes by writing
\bea
Z^{jk}&=&i\left(p^jZ_{\rm T}^k+p^kZ_{\rm T}^j\right)+Z^{jk}_{\rm TT} \label{ipdf-24}\\
Y^{jk}&=&i\left(p^jY^k_{\rm T}+p^kY^j_{\rm T}\right)+Y^{jk}_{\rm TT} \label{ipdf-25}
\tea
where $p_jZ^j_{\rm T}=p_jY^j_{\rm T}=p_jZ^{jk}_{\rm TT}=p_jY^{jk}_{\rm TT}=0$. 

\subsection{The quadratic action}\label{qa}

After separating vector and tensor proper modes, the quadratic terms decouple. For the vectors we get
\bea
S_{NqV}&=&iN\int\;dt\frac{d^3p}{\left(2\pi\right)^3}\;p^2{Y_{\rm T}}_j\left(-p,t\right){Y_{\rm T}}^{j}\left(p,t\right)\label{qa-1}\\
S_{TqV}&=&-\sigma T_0^4 \int\;dt\frac{d^3p}{\left(2\pi\right)^3}\;\left[ \frac43{Y}_{j;0}\left(-p,t\right)V^j\left(p,t\right)
+p^2\alpha {Y}_{j}\left(-p,t\right){Z_{\rm T}}^{j}\left(p,t\right)\right]\label{qa-2}\\
S_{AqV}&=&-\frac12aT_0^5\int\;dt\frac{d^3p}{\left(2\pi\right)^3}\;\left[\frac2{\alpha} p^2 {Y_{\rm T}}_{j;0}\left(-p,t\right){Z_{\rm T}}^{j}
\left(p,t\right)-2p^2{Y_{\rm T}}_{j}\left(-p,t\right)V^{j}\left(p,t\right) \right]\label{qa-3}\\
S_{IqV}&=&\frac{aT_0^5}{\alpha\tau}\int\;dt\frac{d^3p}{\left(2\pi\right)^3}\;p^2 {Y_{\rm T}}_{j} \left(-p,t\right)\;{Z_{\rm T}}^{j}\left(p,t\right)
\label{qa-4}
\tea
and for the tensors
\bea
S_{NqT}&=&\frac{iN}{2}\int\;dt\frac{d^3p}{\left(2\pi\right)^3}\;{Y_{\rm TT}}_{jk}\left(-p,t\right){Y_{\rm TT}}^{jk}\left(p,t\right)\label{qa-5}\\
S_{AqT}&=&-\frac12aT_0^5\int\;dt\frac{d^3p}{\left(2\pi\right)^3}\;\left[\frac1{\alpha}  {Y_{\rm TT}}_{jk;0}\left(-p,t\right){Z_{\rm TT}}^{jk}\left(p,t\right)
\right]\label{qa-6}\\
S_{IqT}&=&\frac{aT_0^5}{2\alpha\tau}\int\;dt\frac{d^3p}{\left(2\pi\right)^3}\; {Y_{\rm TT}}_{jk} \left(-p,t\right)\;{Z_{\rm TT}}^{jk}\left(p,t\right)
\label{qa-7}
\tea

\subsection{The lowest order propagators}\label{lop}

We now return to the conformal fluid case, where the string of fields $\mathcal{X}^K$ may be split into physical vector fields $V^j$ 
and $Z_{\rm T}^j$, auxiliary vector fields $Y^j$ and $Y_{\rm T}^j$, the physical tensor proper field ${Z_{\rm TT}}^{ij}$ and the auxiliary tensor proper field 
${Y_{\rm TT}}^{ij}$. The correlations between a physical and an auxiliary field yield the causal propagators; if the physical field is to the left, 
then it is a retarded propagator. The correlations between physical fields are the symmetric correlations in the theory; the correlations 
between auxiliary fields vanish identically. To lowest order, we obtain decoupled equations for correlations involving only vector fields and those involving only tensor fields.

\subsubsection{Causal vector correlations}\label{cvc}
The causal vector correlations are $\left\langle V^jY^k\right\rangle $ and  $\left\langle {Z_{\rm T}}^jY^k\right\rangle $ on one hand, and 
$\left\langle V^j{Y_{\rm T}}^k\right\rangle $ and  $\left\langle {Z_{\rm T}}^j{Y_{\rm T}}^k\right\rangle $ on the other. These two pairs are decoupled from each 
other. They all have the structure
\be 
\left\langle V^j\left( p,t\right) Y^k\left( q,t'\right) \right\rangle =i\left( 2\pi\right)^3\delta\left( p+q\right) P^{jk}\left[ p\right] 
G_{VY}\left( p,t-t'\right) \label{cvc-1}
\te
where
\be
 P^{jk}\left[ p\right]=\delta^{jk}-\frac{p^jp^k}{p^2} \label{cvc-2}
\te
The equations of motion for the $G_{VY}$, $G_{{Z_{\rm T}}Y}$ pair are
\bea
\sigma T_0^4\left[ \frac43\frac d{dt}G_{VY}-\alpha p^2G_{{Z_{\rm T}}Y}\right] &=&\delta\left( t-t'\right) \label{cvc-3}\\
aT_0^5p^2\left[ G_{VY}+\frac1{\alpha}\frac d{dt}G_{{Z_{\rm T}}Y}\right] +\frac{aT_0^5}{\alpha\tau}p^2G_{{Z_{\rm T}}Y}&=&0
\label{ceq1}
\tea
whose solution is
\bea
G_{VY}&=&\frac{e^{-\left( t-t'\right) /2\tau}}{\sigma T_0^4}\left[ \cos\omega\left( t-t'\right)
+\frac{\sin\omega\left( t-t'\right)}{2\omega\tau} \right] \Theta\left( t-t'\right) \nn
&=& \frac{c_Vp}{\sigma T_0^4\omega}{e^{-\left( t-t'\right) /2\tau}}\cos\left[\;\omega\left( t-t'\right)-\varphi_p\right] 
\Theta\left( t-t'\right) \label{cvc-5}\\
G_{{Z_{\rm T}}Y}&=&\frac{-\alpha}{\omega}\frac{e^{-\left( t-t'\right) /2\tau}}{\sigma T_0^4}\sin\omega\left( t-t'\right)\Theta\left( t-t'\right)
\label{cvc-6}
\tea 
where 
\bea 
\omega&=&\sqrt{\frac34\alpha^2p^2-\frac1{4\tau^2}}\nn
\varphi_p&=&\tan^{-1}\left(\frac1{2\omega\tau}\right)=\frac{\pi}2-\tan^{-1}\left(2\omega\tau\right)
\label{varphip}
\tea 
$c_V=\sqrt{3}\alpha/2$ is the propagation speed for vector modes (cfr. subsection (\ref{incom})), which in the limit we are interested
satisfies $c_V \ll 1$. 

For the second pair $\left\langle V^j{Y_{\rm T}}^k\right\rangle $ and  $\left\langle {Z_{\rm T}}^j{Y_{\rm T}}^k\right\rangle $ we obtain 
\bea
\sigma T_0^4\left[ \frac43\frac d{dt}G_{V{Y_{\rm T}}}-\alpha p^2G_{{Z_{\rm T}}{Y_{\rm T}}}\right] &=&0\nn
aT_0^5p^2\left[ G_{V{Y_{\rm T}}}+\frac1{\alpha}\frac d{dt}G_{{Z_{\rm T}}{Y_{\rm T}}}\right] +\frac{aT_0^5}{\alpha\tau}p^2G_{{Z_{\rm T}}{Y_{\rm T}}}&=&\delta\left( t-t'\right) 
\label{ceq2}
\tea
with solution
\bea
G_{V{Y_{\rm T}}}&=&\frac{3\alpha^2}{4aT_0^5\omega}{e^{-\left( t-t'\right) /2\tau}}\sin\omega\left( t-t'\right)\Theta\left( t-t'\right)
\label{cvc-7} \\
G_{{Z_{\rm T}}{Y_{\rm T}}}&=&\frac{c_V\alpha}{aT_0^5p\omega}{e^{-\left( t-t'\right) /2\tau}}
\cos\left[\; \omega\left( t-t'\right)+\varphi_p \right]  \Theta\left( t-t'\right)  \label{cvc-8}
\tea 

In summary, the causal correlations of vector fields all have the structure

\be 
\left\langle X^j_{\alpha}\left( p,t\right) Y^k_{\beta}\left( q,t'\right) \right\rangle =
i\left( 2\pi\right)^3\delta\left( p+q\right) P^{jk}\left[ p\right]C_{\alpha\beta} e^{-\left( t-t'\right) /2\tau}
\cos \left[\; \omega\left( t-t'\right)-\varphi_{\alpha\beta} \right] \Theta\left( t-t'\right)   \label{cvc-9}
\te
The different correlations are summarized in Table \ref{vectorcausal}.
\begin{table}
	\centering
\begin{tabular}{|c|c|c|c|}\hline
$X_{\alpha}$&$Y_{\beta}$&$C_{\alpha\beta}$&$\varphi_{\alpha\beta} $\\ 
 \hline
$V$& $Y$&$\left( c_Vp\right) /\left( \sigma T_0^4\omega\right) $ & $\varphi_p$ \\ 
\hline
$Z_{\rm T}$& $Y$&$\left( -\alpha\right) /\left( \sigma T_0^4\omega\right) $  &$\pi /2$ \\ 
\hline
$V$& $Y_{\rm T}$&$\left( {3\alpha^2}\right) /\left( {4aT_0^5\omega}\right)$ &$\pi /2$  \\ 
\hline
$Z_{\rm T}$& $Y_{\rm T}$&$\left( c_V\alpha\right) \left( aT_0^5p\omega\right) $ &$-\varphi_p$ \\ 
\hline
\end{tabular}
\caption{Summary of vector causal propagators. $\omega$ and $\varphi_p$ are defined in eq. (\ref{varphip}).
\label{vectorcausal}}
\end{table}

\subsubsection{Causal tensor correlations}\label{sctc}
The only causal tensor correlation is 
\be 
\left\langle Z_{\rm TT}^{ij}\left( p,t\right) {Y_{\rm TT}}_{kl}\left( q,t'\right) \right\rangle =
i\left( 2\pi\right)^3\delta\left( p+q\right) {{S_{\rm TT}}^{ij}}_{kl}\left[ p\right] G_{{Z_{\rm TT}}{Y_{\rm TT}}}\left( p,t-t'\right)
\label{octc} 
\te
where
\be 
{{S_{\rm TT}}^{ij}}_{kl}\left[ p\right]=\frac1{2}\left[ P^i_kP^j_l+P^i_lP^j_k-P^{ij}P_{kl}\right] 
\label{Sijkl}
\te 
It obeys the equation 
\be 
\frac{aT_0^5}{2\alpha}\left[ \frac d{dt}+\frac1{\tau}\right] G_{Z_{\rm TT}Y_{\rm TT}}= \delta\left( t-t'\right) 
\label{ctcem}
\te
with solution 
\be 
G_{Z_{\rm TT}Y_{\rm TT}}=\frac{2\alpha}{aT_0^5}e^{-\left( t-t'\right) /\tau}\Theta\left(t-t'\right)
\label{ctc}
\te 
Observe that the tensor modes do not propagate, and their momentum dependence is trivial.

\subsubsection{Hadamard vector correlations}\label{Hvc}

The vector correlations are $\left\langle V^jV^k\right\rangle $, 
$\left\langle V^jZ_{\rm T}^k\right\rangle $ and 
$\left\langle Z_{\rm T}^jZ_{\rm T}^k\right\rangle $. They have the structure 
\be 
\left\langle V^j\left( p,t\right) V^k\left( q,t'\right) \right\rangle =\left( 2\pi\right)^3\delta\left( p+q\right) P^{jk}\left[ p\right] 
G_{1VV}\left( p,t-t'\right) \label{Hvc-1}
\te
They may be derived from eq. (\ref{correlations}). For example 
\be 
G_{1VV}\left( p,t-t'\right) =2Np^2\int^{\mathrm{min}\left( t,t'\right) }dt''\;G_{VY_T}\left( p,t-t''\right) G_{VY_T}\left( p,t'-t''\right) 
\label{Hvc-2}
\te
Explicitly, 
\bea 
G_{1VV}\left( p,t-t'\right) &=&\frac 3{4\sigma T_0^3}e^{-|t-t'| /2\tau}\left[ \cos\omega|t-t'|
+\frac{\sin\omega|t-t'|}{2\omega\tau}\right]\nn
&=&\frac{3c_Vp}{4\sigma T_0^3\omega}e^{-|t-t'| /2\tau}\cos\left[\; \omega|t-t'|-\varphi_p\right] \label{Hvc-3}
\tea
Similarly 
\be 
G_{1VZ_{\rm T}}\left( p,t-t'\right) =2Np^2\int^{\mathrm{min}\left( t,t'\right) }dt''\;G_{VY_{\rm T}}\left( p,t-t''\right) G_{Z_{\rm T}Y_{\rm T}}\left( p,t'-t''\right) 
\label{Hvc-4}
\te
Now, from the equations for the propagators 
\be 
G_{1VZ_{\rm T}}\left( p,t-t'\right) =2Np^2\int^{\mathrm{min}\left( t,t'\right) }dt''\;G_{VY_{\rm T}}\left( p,t-t''\right) \frac4{3\alpha p^2}
\frac d{dt'}G_{VY_{\rm T}}\left( p,t'-t''\right) \label{Hvc-5}
\te 
Since the integrand vanishes at the upper limit, we find
\bea 
G_{1VZ_{\rm T}}\left( p,t-t'\right) &=& \frac4{3\alpha p^2}\frac d{dt'}G_{1VV}\left( p,t-t'\right)\nn
&=&\frac {3\alpha}{4\sigma T_0^3\omega}e^{-|t-t'| /2\tau}{\sin\omega|t-t'|}\mathrm{sign}\left( t-t'\right) 
\label{Hvc-6}
\tea
The correlation $G_{1Z_{\rm T}Z_{\rm T}}$ is even in $t-t'$, and so there is no loss of generality in assuming $t>t'$. In this case, we get 
\bea 
G_{1Z_{\rm T}Z_{\rm T}}\left( p,t-t'\right) &=& \frac4{3\alpha p^2}\frac d{dt}G_{1VZ_{\rm T}}\left( p,t-t'\right)\nn
&=&\frac {1}{\sigma T_0^3p^2}e^{-|t-t'| /2\tau}\left[ \cos\omega|t-t'|-\frac{\sin\omega|t-t'|}{2\omega\tau}\right] \nn
&=&\frac{c_V}{\sigma T_0^3\omega p}e^{-|t-t'| /2\tau} \cos\left[\;\omega|t-t'|+\varphi_p\right]
\label{Hvc-7}
\tea
In summary, the structure of vector correlations is 
\be 
\left\langle X^j_{\alpha}\left( p,t\right) X^k_{\beta}\left( q,t'\right) \right\rangle =
\left( 2\pi\right)^3\delta\left( p+q\right) P^{jk}\left[ p\right]C_{1\alpha\beta} e^{-| t-t'| /2\tau}
\cos \left[\; \omega| t-t'|-\varphi_{1\alpha\beta} \right] \left( \mathrm{sign}\left( t-t'\right) \right) ^{q_{\alpha\beta}}
\label{Hvc-8}
\te
They are summarized in Table \ref{vectorcorr}
\begin{table}
		\centering
\begin{tabular}{|c|c|c|c|c|}\hline
$X_{\alpha}$&$X_{\beta}$&$C_{1\alpha\beta}$&$\varphi_{1\alpha\beta} $& $q_{\alpha\beta}$\\ 
 \hline
$V$& $V$&$\left( {3c_Vp}\right) /\left( {4\sigma T_0^3\omega}\right)  $ & $\varphi_p$ & $0$\\ 
\hline
$V$& $Z_{\rm T}$&$\left( {3\alpha}\right)/ \left( {4\sigma T_0^3\omega}\right)  $  &$\pi /2$ & $1$ \\ 
\hline
$Z_{\rm T}$& $Z_{\rm T}$&$\left( {c_V}\right) /\left( {\sigma T_0^3\omega p}\right) $ &$-\varphi_p$& $0$  \\ 
\hline
\end{tabular}
\caption{Summary of vector correlations. $\varphi_p$ is defined in eq. (\ref{varphip})}
\label{vectorcorr}
\end{table}

\subsubsection{Hadamard tensor correlations}\label{tc}
The remaining correlation is trivial. It has the structure 

\be 
\left\langle {Z_{\rm TT}}^{jk}\left( p,t\right) {Z_{\rm TT}}^{lm}\left( q,t'\right) \right\rangle =\left( 2\pi\right)^3\delta\left( p+q\right){S_{\rm TT}}^{jklm}
\left[ p\right] G_{1Z_{\rm TT}Z_{\rm TT}}\left( p,t-t'\right) 
\te
where the projector $S_{\rm TT}$ is defined in eq. (\ref{Sijkl}), and
\bea
G_{1{Z_{\rm TT}}{Z_{\rm TT}}}&=&N\left(\frac{2\alpha}{aT_0^5}\right)^2\frac{\tau}2e^{-|t-t'| /\tau}\nn
&=&\frac{2}{\sigma T_0^3} e^{-|t-t'| /\tau}
\label{Hadatc}
\tea
We may check that in the limit $\tau\to 0$ we recover Landau-Lifshitz theory. Indeed in this limit we may approximate 
\be
G_{1{Z_{\rm TT}}{Z_{\rm TT}}}=\frac{4\tau}{\sigma T_0^3} \delta\left( t-t'\right) 
\label{HadatcLL}
\te
In this limit, the self correlation for the tensor proper part of the viscous EMT (cfr. eq. (\ref{current-emt})) is
\bea 
\left\langle {\Pi_{\rm TT}}^{jk}\left( p,t\right) {\Pi_{\rm TT}}^{lm}\left( q,t'\right) \right\rangle &=&\alpha^2\sigma^2T_0^8\left\langle {Z_{\rm TT}}^{jk}\left( p,t\right) {Z_{\rm TT}}^{lm}\left( q,t'\right) \right\rangle \nn
&=&\left( 2\pi\right)^3\delta\left( p+q\right){S_{\rm TT}}^{jklm}\left[ p\right]\,4\,T_0\,\eta\,\delta\left( t-t'\right) 
\label{LLhf}
\tea
where we have used eq. (\ref{alfasq}). We thus recover the Landau-Lifshitz result \cite{LLflucthyd,LLsmII} in this limit.

\section{Nonlinear fluctuations around equilibrium}\label{Nlfae}

In this work we shall show the application of the field theory technics to a consistent causal theory. We choose to focus on the interaction between tensor and vector modes. In this section we compute the one-loop
corrections to the tensor
propagators found above, and apply the results to derive the 
corresponding
fluctuations in the energy and entropy densities. 
To this purpose we need the cubic terms in the ``classical''
action eq. (\ref{class2}).

Considering heavy-ion collisions, we may ask at which stage of the fireball evolution are the loop corrections significative. We shall show below (eq. (\ref{p_t})) that this holds for $p > p_T\sim T_0\left( T_0\tau\right)^{-3/8}$. Since we work within the free-streaming approximation, it implies that the spatial correlation at equal times is non-trivial for distances $r<p_T^{-1}<c_V\tau<\tau$. These are the relevant scales at the very early stages of a heavy-ion collision.

\subsection{Cubic terms in the ``classical'' action}\label{ctca}

We shall not need the cubic terms which do not contain tensor modes. For those which contain tensor fields, we distinguish

a) Terms that only contain ${Y_{\rm TT}}_{jk}\left(-p,t\right)$:
These terms naturally split into two
\be 
S_{{Y_{\rm TT}}}=S_{{Y_{\rm TT}}VV}+S_{{Y_{\rm TT}}V{Z_{\rm T}}}
\label{ctca-1}
\te 
\bea 
S_{{Y_{\rm TT}}VV}&=&- \frac{5aT_0^5}{2\alpha\tau}\int\;dt\frac{d^3p}{\left(2\pi\right)^3}\frac{d^3q}{\left(2\pi\right)^3}\;D_{jkrs}\left( p,q\right)  {Y_{\rm TT}}^{jk}\left(-p,t\right) V^r\left(p-q,t\right)V^{s}\left(q,t\right)\nn
S_{{Y_{\rm TT}}V{Z_{\rm T}}}&=&- \frac {aT_0^5}{\alpha}\int\;dt\frac{d^3p}{\left(2\pi\right)^3}\frac{d^3q}{\left(2\pi\right)^3}\;E_{jkrs}\left( p,q\right)  {Y_{\rm TT}}^{jk}\left(-p,t\right) V^r\left(p-q,t\right){Z_{\rm T}}^{s}\left(q,t\right)
\label{ctca-2}
\tea 
where
\bea 
D_{jkrs}\left( p,q\right) &=&\delta_{rk}\delta_{js}\nn
E_{jkrs}\left( p,q\right) &=& \left( q^2+p_lq^l\right)\delta_{sj}\delta_{kr}+q^k\left(\delta_{js}p_r+p_s\delta_{jr}\right)
\label{ctca-3}
\tea

b) Terms that only contain ${Z_{\rm TT}}_{lm}\left(q,t\right)$: 
These terms also split as
\be 
S_{Z_{\rm TT}}=S_{{Y_{\rm T}}V{Z_{\rm TT}}}+S_{Y{Z_{\rm T}}{Z_{\rm TT}}}
\te 
\bea 
S_{{Y_{\rm T}}V{Z_{\rm TT}}}&=&\frac {aT_0^5}{\alpha}\int\;dt\frac{d^3p}{\left(2\pi\right)^3}\frac{d^3q}{\left(2\pi\right)^3}\;F_{rslm}\left( p,q\right)  {Y_{\rm T}}^{r}\left(-p,t\right) V^s\left(p-q,t\right){Z_{\rm TT}}^{lm}\left(q,t\right)\nn
S_{Y{Z_{\rm T}}{Z_{\rm TT}}}&=&- \sigma T_0^4\int\;dt\frac{d^3p}{\left(2\pi\right)^3}\frac{d^3q}{\left(2\pi\right)^3}\;G_{rslm}\left( p,q\right)  Y^{r}\left(-p,t\right) {Z_{\rm T}}^s\left(p-q,t\right){Z_{\rm TT}}^{lm}\left(q,t\right)
\label{ctca-4}
\tea 
where
\bea 
F_{rslm}\left( p,q\right) &=&\frac34\alpha^2p^2\delta_{mr}\delta_{ls}+2   p^l\delta_{mr} p_s+  p^m p_l\delta_{rs} \nn
G_{rslm}\left( p,q\right) &=&  p_m \left(\left(p-q\right)^r\delta_{ls}+\left(p-q\right)^l\delta_{rs}\right)+\delta_{rm}\left(p_s\left( p-q\right) ^l+p_k\left( p-q\right)^k\delta_{sl}\right)
\label{ctca-5}
\tea

We shall not need the explicit form of the remaining terms:

c) terms that contain both ${Y_{\rm TT}}_{jk}\left(-p,t\right)$ and ${Z_{\rm TT}}_{lm}\left(q,t\right)$,

d) terms quadratic in  ${Z_{\rm TT}}_{lm}\left(q,t\right)$.

\subsection{Tensor self-energy}\label{Tse}

We now turn to the derivation of the self energy for tensor modes
\be
\Sigma_{jklm}\left(-p,q,t-t'\right)=2\frac{\delta\Gamma_Q}{\delta \left\langle {Y_{\rm TT}}^{jk}\left(-p,t\right){Z_{\rm TT}}^{lm}\left(q,t'\right)\right\rangle} \label{ctca-6}
\te
We may split the self energy into two contributions, one with only vector propagators in internal lines, and the other with one vector and one tensor modes. Our goal is to derive the momentum dependence of the self-energy and the noise kernel. Now, because the lowest order tensor propagators are momentum independent, the Feynman graphs containing them are momentum independent too. For this reason we shall not compute them. 

The self-energy, considering only the Feynman graphs with vector propagators in internal lines, is
\be
\Sigma^{\left( V\right) }_{jklm}\left(-p,q,t-t'\right)=i\left\langle\frac{\delta \left( S_{{Y_{\rm TT}}VV}+ S_{{Y_{\rm TT}}V{Z_{\rm T}}}\right) }{\delta {Y_{\rm TT}}^{jk}\left(-p,t\right)}\frac{\delta\left(S_{{Y_{\rm T}}V{Z_{\rm TT}}}+S_{Y{Z_{\rm T}}{Z_{\rm TT}}} \right) }{\delta {Z_{\rm TT}}^{lm}\left(q,t'\right)} \right\rangle 
\label{ctca-7}
\te
It is understood that after computing the functional derivatives one must project back onto tensor proper modes. The terms involving $S_{{Y_{\rm TT}}VV}$ are suppressed by one power of $\tau$ and will not be computed. Let us begin with
\bea
&&\Sigma^{\left( V1\right) }_{jklm}\left(-p,q,t-t'\right)=i\left\langle\frac{\delta S_{{Y_{\rm TT}}V{Z_{\rm T}}}}{\delta {Y_{\rm TT}}^{jk}\left(-p,t\right)}\frac{\delta S_{{Y_{\rm T}}V{Z_{\rm TT}}} }{\delta {Z_{\rm TT}}^{lm}\left(q,t'\right)} \right\rangle \nn
&=&-i \frac {a^2T_0^{10}}{\alpha^2}\int\;\frac{d^3q'}{\left(2\pi\right)^6}\frac{d^3p'}{\left(2\pi\right)^6}\;E_{jkrs}\left( p,q'\right) F_{uvlm}\left( p',q\right)\nn 
&&\left\{\left\langle  V^r\left(p-q',t\right){Y_{\rm T}}^{u}\left(-p',t'\right)\right\rangle\left\langle {Z_{\rm T}}^{s}\left(q',t\right)   V^v\left(p'-q,t'\right)\right\rangle\right.\nn
&+&\left.\left\langle  V^r\left(p-q',t\right)V^v\left(p'-q,t'\right)\right\rangle\left\langle {Z_{\rm T}}^{s}\left(q',t\right) {Y_{\rm T}}^{u}\left(-p',t'\right)  \right\rangle\right\}
\label{ctca-8}
\tea
Observe that $\Sigma^{\left( V1\right) }_{jklm}$ vanishes unless $t\ge t'$, which we assume. Moreover, the graph in the second line of the right hand side vanishes when $t=t'$ and we shall not compute it. We then have
\be
\Sigma^{\left( V1\right) }_{jklm}\left(-p,q,t-t'\right)= \frac {9\alpha aT_0^{2}}{16\sigma }\delta\left( p-q\right)(2\pi)^{-6} e^{-( t-t') /\tau}\Theta(t-t')\sigma^{\left( V1\right) }_{jklm}
\label{ctca-9}
\te
where 
\bea
&&\sigma^{\left( V1\right) }_{jklm}=\int\;{d^3q'}\;E_{jkrs}\left( p,q'\right) F_{uvlm}\left( q',p\right)P^{rv}\left[ p-q'\right]P^{su}\left[ q'\right]\nn 
&& \frac{\left|p-q'\right|}{q'\omega_{q'}\omega_{p-q'}}\cos \left[\; \omega_{p-q'}\left( t-t'\right)-\varphi_{p-q'} \right] \cos \left[\; \omega_{q'}\left( t-t'\right)+\varphi_{q'} \right] 
\label{ctca-10}
\tea
On dimensional grounds, we see that the tensor self energy has 
units of $T^4$ as it should. $\sigma^{\left( V1\right) }_{jklm}$ 
has units of $p^5$.

To obtain the true self energy we must project back on the transverse components, symmetric and traceless in $jk$ and $lm$. If $p$ lies in the $z$ direction, this means we only need $\sigma^{\left( V1\right) }_{abcd}$, where the indices run from $1$ to $2$. Since there are no preferred directions, we will obtain
\be 
\sigma^{\left( V1\right) }_{abcd}=A\delta_{ac}\delta_{bd}+B\delta_{ad}\delta_{bc}+C\delta_{ab}\delta_{cd}
\label{ctca-11}
\te 
Symmetrization on $ab$ yields
\be 
\sigma^{\left( V1\right) }_{abcd}\to \frac12\left(A+B\right)\left(\delta_{ac}\delta_{bd}+\delta_{ad}\delta_{bc}\right)+C\delta_{ab}\delta_{cd} \label{ctca-12}
\te 
and removing the trace on $ab$ we get
\be 
\sigma^{\left( V1\right) }_{abcd}\to \left(A+B\right){S_{\rm TT}}_{abcd}\frac12\left(\delta_{ac}\delta_{bd}+\delta_{ad}\delta_{bc}-\delta_{ab}\delta_{cd}\right) \label{ctca-13}
\te 
where ${S_{\rm TT}}_{abcd}$ is the restriction to the case where $p$ is on the third direction of  the projector eq. (\ref{Sijkl}). This means that the physical self energy takes the form
\be
\Sigma^{\left( V1\right) }_{ijlm}\left(-p,q,t-t'\right)=\frac {9\alpha aT_0^{2}}{16\sigma }\delta\left( p-q\right)(2\pi)^{-6} e^{-| t-t'| /\tau}\Theta(t-t')\sigma^{\left( V1\right) }_{phys}{S_{\rm TT}}_{ijlm} \label{ctca-14}
\te 
where 
\be
\sigma^{\left( V1\right) }_{phys}=\frac12\int\;{d^3q'}\;W\left[p,q'\right]\frac{\left|p-q'\right|}{\omega_{p-q'}}\cos \left[\; \omega_{p-q'}\left( t-t'\right)-\varphi_{p-q'} \right] \cos \left[\; \omega_{q'}\left( t-t'\right)+\varphi_{q'} \right]  
\label{ctca-15}
\te
with
\be
W\left[p,q'\right]=\frac{1}{q'\omega_{q'}}{S_{\rm TT}}^{ijkl}\left[p\right]E_{jkrs}\left( p,q'\right) F_{uvlm}\left( q',p\right)P^{rv}\left[ p-q'\right]P^{su}\left[ q'\right] \label{ctca-16}
\te
observe that $W$ has units of $p^2$.

To proceed, we shall make an important simplification. As we have seen, the ``classical'' equations for the correlations eqs. (\ref{ceq1}, \ref{ceq2}, \ref{ctcem}) are local in time. We assume that the main loop corrections to these equations are those that are local in time too. Therefore we shall seek only the singular terms in the self energy and the noise kernel, namely the terms which are proportional to $\delta\left( t-t'\right) $. Since the propagators themselves are regular functions, any such singular term can only result from the asymptotic large $\vert q'\vert$ region of the integration domain. 
In this region $\omega_{q'}$ and $\omega_{p-q'}$ are both real. In principle different schemes, such as BRSSS \cite{romat17,brsss}, DNMR \cite{dnmr}, Anisotropic Hydrodynamics \cite{Stri14} or DTTs, may lead to different UV behaviors and, moreover, different procedures can be adopted to regularize the large momentum contribution to the Feynman graphs. Therefore, the ensuing discussion is predicated on our choice of DTTs and dimensional regularization.

To compute $\sigma^{\left( V1\right) }_{phys}$ we keep only the terms which are slow and not zero in the coincidence limit

\bea
&&\cos \left[\; \omega_{p-q'}\left( t-t'\right)-\varphi_{p-q'} \right] \cos \left[\; \omega_{q'}\left( t-t'\right)+\varphi_{q'} \right]  \nn
&&\simeq\frac12\cos\left[\left(\omega_{p-q'}-\omega_{q'}\right)\left( t-t'\right)\right]\cos\left[\varphi_{p-q'}\right]\cos\left[\varphi_{q'}\right] \label{ctca-17}
\tea
So
\be
\sigma^{\left( V1\right) }_{phys}\approx\frac14\int\;{d^3q'}\;W\left[p,q'\right]\frac{\left|p-q'\right|}{\omega_{p-q'}}\cos \left[\; \left(\omega_{p-q'}-\omega_{q'}\right)\left( t-t'\right)\right]\cos\left[\varphi_{p-q'}\right]\cos\left[\varphi_{q'}\right]
\label{ctca-18}
\te
Assume again the $p^j=\left(0,0,p\right)$. Then $\omega_{p-q'}-\omega_{q'}$ vanishes when $q'=\left(q_T,p/2\right)$, with $q_T$ in the $x,y$ plane. We write $q'=\left(q_T,p/2+\delta q_3\right)$ to get (where explicit, $q'=\left(q_T,p/2\right)$)
\be
\omega_{p-q'}-\omega_{q'}\approx-\frac{c_V^2p}{\omega_{q'}}\delta q_3
\label{ctca-19}
\te
Evaluating the prefactors at $\delta q_3=0$ we may integrate over 
$\delta q_3$ to obtain the singular part of the self-energy
\be
\sigma^{\left( V1\right) }_{phys}\approx\frac{\pi}{2\,c_V^2}\delta\left( t-t'\right)\int\;{d^2q_T}\;W\left[p,q'\right]\frac{q'}{p}\cos^2 \left[\varphi_{q'} \right]  \label{ctca-20}
\te
where, in the free-streaming limit $\tau\rightarrow\infty$,
\be
\cos^2 \left[\varphi_{q'} \right] =\frac{1+\cos \left[2\varphi_{q'} \right]}2=1-\frac1{4\,\omega_{q'}^2\tau^2}\approx 1
\label{ctca-21}
\te
The remaining task is to compute (we neglect terms proportional to 
$\alpha^2$)
\bea
&&J^{(V1)}=\frac{\pi}{2\,c_V^2}\int\;{d^2q_T}\;W\left[p,q'\right]\frac{q'}{p}\nn
&=&\frac{\pi}{2\,p\,c_V^3}\int\;\frac{d^2q_T\;\left[-q_{T}^{8}+\frac58\,q_T^6\,p^2-\frac18\,q_T^4\,p^4+\frac{17}{128}\,q_T^2\,p^6\right]}{\left(q_T^2+\frac{p^2}{4}\right)^2\left(q_T^2+\frac{p^2}{4}-\frac1 {4c_V^2\tau^2}\right)^{1/2}}
\label{jv1}
\tea

We have integrals of the form
\bea
s^{(n)}(p)&=&\int \frac{d^2q_T\left(q_T^2\right)^n}{\left(q_T^2+\frac{p^2}{4}\right)^{2}\left(q_T^2+\frac{p^2}{4}-\frac1 {4c_V^2\tau^2}\right)^{1/2}}\nn
&=&\frac{\Gamma[5/2]}{\Gamma[2]\Gamma[1/2]}\int_0^1dx\,(1-x)x^{-1/2}\int\frac{d^2q_T\,q_T^{2n}}{\left(q_T^2+\frac{p^2}{4}-\frac{x}{4c_V^2\tau^2}\right)^{5/2}} \label{ctca-22}
\tea
We compute them in the scheme of dimensional regularization in 
$D=2-\epsilon$ dimensions. Following \cite{ramond,kleinert} we get
\bea
s^{n}(p)&=&\pi^{1-\epsilon/2}\frac{\Gamma[1-\epsilon/2+n]\Gamma[3/2+\epsilon/2-n]}{\Gamma[2]\Gamma[1/2]\Gamma\left[1-\epsilon/2\right]}\nn
&&\mu^{\epsilon}\int_0^1dx\,(1-x)x^{-1/2}\left[\frac14\left(p^2-\frac{x}{c_V^2\tau^2}\right)\right]^{n-\epsilon/2-3/2}.
\label{ctca-23}
\tea
As we see the limit $\epsilon\rightarrow0$ is well defined for any integer $n$, so we take $\epsilon=0$ ($D=2$) straightforwardly. Further in the free-streaming limit $\tau\rightarrow\infty$, we get
\be
J^{(V1)}=\frac{493\,\pi^2}{480}\frac{p^4}{c_V^3} \label{ctca-24}
\te
\be
\Sigma^{\left( V1\right) }_{ijlm}\left(-p,q,t-t'\right)=\frac{493\sqrt{3}\,\pi^2}{2560}\,\frac{1}{(2\pi)^6}\frac { a\,T_0^{2}\,p^4}{c_V^2\,\sigma }\delta\left( p-q\right) \delta(t-t')\,{S_{\rm TT}}_{ijlm}(p) \label{ctca-25}
\te 
We now consider the other graph
\be
\Sigma^{\left( V2\right) }_{jklm}\left(-p,q,t-t'\right)=i\left\langle\frac{\delta  S_{{Y_{\rm TT}}V{Z_{\rm T}}} }{\delta {Y_{\rm TT}}^{jk}\left(-p,t\right)}\frac{\delta S_{Y{Z_{\rm T}}{Z_{\rm TT}}}  }{\delta {Z_{\rm TT}}^{lm}\left(q,t'\right)} \right\rangle
\label{ctca-26}
\te
Repeating the same steps as in the previous case we get 
\be
\Sigma^{\left( V2\right) }_{jklm}\left(-p,q,t-t'\right)=-\frac{aT_0^2c_V^2e^{-(t-t')/\tau}}{(2\pi)^{6}\alpha\sigma }\Theta(t-t')\delta(p-q)\sigma_{phys}^{(V2)}(p,t-t')\,{S_{\rm TT}}_{jklm}(p).
\label{ctca-27}
\te

To compute $\sigma^{\left( V2\right) }_{phys}$ we keep only the terms which are slow and not zero in the coincidence limit
\be
\sigma_{phys}^{(V2)}(p,t-t')\simeq\frac{\pi}{2\,c_V^2} \delta(t-t')\int \frac{d^2k_{\perp}}{p\,\omega_k}\,\frac{\left[ \frac12k_{\perp}^8+k_{\perp}^6p^2+\frac9{32}k_{\perp}^4p^4+\frac{9}{64}k_{\perp}^2p^6-\frac3{128}p^8\right] }{\left(k_{\perp}^2+\frac{p^2}4\right)^2}
\label{ctca-28}
\te
Finally, in the free-streaming limit, we get
\be
\sigma_{phys}^{(V2)}(p,t-t')=\frac{17\pi^2}{60}\frac{p^4}{c_V^3}\delta(t-t') \label{ctca-29}
\te
and
\be
\Sigma^{\left( V2\right) }_{jklm}\left(-p,q,t-t'\right)
=-\frac{17\sqrt{3}}{240}\frac{\pi^2}{(2\pi)^{6}}\frac { a\,T_0^{2}\,p^4}{c_V^2\,\sigma }\delta\left( p-q\right) \delta(t-t')\,{S_{\rm TT}}_{jklm}(p) \label{ctca-30}
\te

The total physical self-energy induced by vector fluctuations reads, in the free-streaming limit,
\be
\Sigma^{\left( V\right) }_{jklm}\left(-p,q,t-t'\right)=\frac{\gamma_{\Sigma}}{(2\pi)^{3}}\frac {\sqrt{3}\, a\,T_0^{2}\,p^4}{4\,c_V^2\,\sigma }\delta\left( p-q\right) \delta(t-t')\,{S_{\rm TT}}_{jklm}(p)\label{ctca-31}
\te
where
\be
\gamma_{\Sigma}=\frac{187}{384}\frac{\pi^2}{(2\pi)^{3}}
\label{ctca-32}
\te

\subsection{Tensor noise-kernel}\label{Tnk}

The derivation of the noise kernel begins with
\be
N_{jklm}\left(-p,q,t-t'\right)=-2i\frac{\delta\Gamma_Q}{\delta \left\langle {Y_{\rm TT}}^{jk}\left(-p,t\right){Y_{\rm TT}}^{lm}\left(q,t'\right)\right\rangle} \label{Tnk-1}
\te
As before, we may split the noise kernel into two contributions, one with only vector propagators in internal lines, and the other with one vector and one tensor propagator. Since the latter is momentum independent, we shall compute the former only:
\be
N^{\left( V\right) }_{jklm}\left(-p,q,t-t'\right)=\left\langle\frac{\delta \left( S_{{Y_{\rm TT}}VV}+ S_{{Y_{\rm TT}}V{Z_{\rm T}}}\right) }{\delta {Y_{\rm TT}}^{jk}\left(-p,t\right)}\frac{\delta\left( S_{{Y_{\rm TT}}VV}+ S_{{Y_{\rm TT}}V{Z_{\rm T}}}\right) }{\delta {Y_{\rm TT}}^{lm}\left(q,t'\right)} \right\rangle 
\label{Tnk-2}
\te
It is understood that after computing the functional derivatives one must project back onto tensor proper modes.

The terms involving $S_{{Y_{\rm TT}}VV}$ are suppressed by one power of $\tau$ and will not be computed. In consequence the noise kernel reads
\be
N^{\left( V\right) }_{jklm}\left(-p,q,t-t'\right)=\left\langle\frac{\delta S_{{Y_{\rm TT}}V{Z_{\rm T}}} }{\delta {Y_{\rm TT}}^{jk}\left(-p,t\right)}\frac{\delta S_{{Y_{\rm TT}}V{Z_{\rm T}}} }{\delta {Y_{\rm TT}}^{lm}\left(q,t'\right)} \right\rangle \label{Tnk-3}
\te
As before, we seek the singular part of the noise kernel. The calculation follows the same steps as the self-energy computation.
We get  
\be
N^{\left( V\right) }_{jklm}\left(-p,q,t-t'\right)
=\left(\frac{a\,T_0^5}{\alpha}\right)^2\frac{3\,c_V^2e^{-|t-t'|/\tau}}{4(2\pi)^{6}T_0^6\sigma^2}\delta(p-q)\,{S_{\rm TT}}_{jklm}(p)\,n_{phys}^{(V)} \label{Tnk-4}
\te
To compute $n^{\left( V\right) }_{phys}$ we keep only the terms which are slow and not zero in the coincidence limit
\be
n_{phys}^{(V)}\simeq\frac{\pi}{2\,c_V^2}\delta(t-t')\int \frac{d^2k_{\perp}}{p\,\omega_{k}}\frac{\left[ \frac12k_{\perp}^8+\frac94k_{\perp}^6p^2+\frac{45}{32}k_{\perp}^4p^4+\frac{15}{32}k_{\perp}^2p^6+\frac9{128}p^8\right]}{\left(k_{\perp}^2+\frac{p^2}4\right)^2} \label{Tnk-5}
\te
Finally, in the free-streaming limit, we get
\be
n_{phys}^{(V)}(p,t-t')=\frac{73\,\pi^2}{240}\,\frac{p^4}{c_V^3}\,\delta(t-t') \label{Tnk-6}
\te
and
\be
N^{\left( V\right) }_{jklm}\left(-p,q,t-t'\right)
=\frac{\gamma_N}{(2\pi)^3}\frac{3\,a^2\,T_0^4\,p^4}{4\,c_V^3\,\sigma^2}\,\delta(p-q)\,\delta(t-t')\,{S_{\rm TT}}_{jklm}(p)
\label{Tnk-7}
\te
where
\be 
\gamma_N= \frac{73\,\pi^2}{320\,(2\pi)^3} \label{Tnk-8}
\te

\subsection{Nonlinear tensor correlations}\label{Nltc}
Adding the singular self-energy term to the classical equation (\ref{ctcem}) we get the one-loop corrected equation for the causal
tensor propagator
\be
\frac{aT_0^5}{2\alpha}\left[ \frac d{dt}+\frac1{\tau_1}\right] G_{{Z_{\rm TT}}{Y_{\rm TT}}}= \delta\left( t-t'\right)
\label{ctcem1l}
\te
where
\be
\frac1{\tau_1}=\frac1{\tau}\left[1+{\gamma_{\Sigma}}\left(\frac p{ p_{\rm T} }\right)^4\right] \label{Tnk-9}
\te
and
\be
p_{\rm T}=\left(\frac{c_V\sigma T_0^3}{\tau}\right)^{1/4}=\left(\frac{p_{\rm L}^3}{c_V\,\tau}\right)^{1/4},\label{p_t}
\te
with $p_{\rm L}$ in (\ref{ploop}), being the largest value of $p$ that makes the loop expansion consistent. Since $c_V\tau\sim \sqrt{\tau/T}$, in the free-streaming limit ($\tau\to\infty$) we get $p_{\rm T}\ll p_{\rm L}$.

Therefore the causal correlation becomes (cfr. eq. (\ref{ctc}))
\be
G_{{Z_{\rm TT}}{Y_{\rm TT}}}=\frac{2\alpha}{aT_0^5}e^{-\left( t-t'\right) /\tau_1}\Theta\left(t-t'\right)
\label{ctc1l},
\te
and the symmetric tensor correlation, analogous to the eq. (\ref{Hadatc}), reads
\bea
G_{1{{Z_{\rm TT}}{Z_{\rm TT}}}}&=&\left[N+{\gamma_N}\frac{3\,a^2\,T_0^4\,p^4}{4c_V^3\,\sigma^2}\right]\left(\frac{2\alpha}{aT_0^5}\right)^2\frac{\tau_1}2e^{-|t-t'| /\tau_1}\nn
&=&\frac{2}{\sigma T_0^3} e^{-|t-t'| /\tau_1}\frac{\tau_1}{\tau}\left[1+{\gamma_N} \left(\frac p{ p_{\rm T} }\right)^4 \right].
\label{Hadatc1l}
\tea
Within the free-streaming approximation, we have two ranges of interest $p\ll p_{\rm T}$ and $p_{\rm T}\ll p<p_{\rm L}$. For the former we recover the spatially flat spectrum in (\ref{Hadatc}) and for the latter we obtain
\bea
G_{1Z_{\rm TT}Z_{\rm TT}}&=&\frac{4\tau}{\sigma T_0^3}\left(\frac{\tau_1}{\tau}\right)^2\left[1+{\gamma_N} \left(\frac p{ p_{\rm T} }\right)^4 \right]\delta(t-t')\\
&=&\frac{4\,\gamma_N}{\gamma_{\Sigma}^2}\frac{c_V}{p^4} \delta\left( t-t'\right).
\label{HadatcLL1lUV}
\tea

The previous analysis clearly shows a transition from a flat spectrum for $p\ll p_{\rm T}$ to a power law spectrum $p^{-4}$ for $p_{\rm T}\ll p< p_{\rm L}$.

\subsection{Entropy and EMT fluctuations}\label{EEMTf}
Having found the tensor correlations in the free-streaming limit 
($\tau\to\infty$), we can easily derive the fluctuations in the viscous EMT, the energy density and the entropy density. 
Since for large wavelengths we recover the zeroth order propagators and therefore the Landau-Lifshitz expression for the fluctuations of $\Pi^{\mu\nu}_{\rm TT}$, cfr. eq. (\ref{LLhf}), we only quote the expression for the viscous EMT for $p_{\rm T}\ll p< p_{\rm L}$, which reads
\be
\left\langle {\Pi_{\rm TT}}^{jk}\left( p,t\right) {\Pi_{\rm TT}}^{lm}\left( q,t'\right) \right\rangle=\left( 2\pi\right)^3\frac{16\,\gamma_N}{3\,\gamma_{\Sigma}^2}\frac{c_V^3\sigma^2T_0^8 }{p^4}\delta\left( p+q\right){S_{\rm TT}}^{jklm}\left[ p\right]\delta\left( t-t'\right) \label{EEMTf-1}
\te
The contribution to the energy density in this model is
\be
\rho= \sigma T_0^4\left[1+\frac74{Z_{\rm TT}}^{jk}{Z_{\rm TT}}_{jk}\right]. \label{EEMTf-2}
\te
To compute the expectation value of ${Z_{\rm TT}}^{jk}{Z_{\rm TT}}_{jk}$ in the equal time limit we use the full form eq. (\ref{Hadatc1l}) to get
\be
\left\langle \rho\right\rangle= \sigma T_0^4\left[1+\frac{7}{\sigma T_0^3}\int\frac{d^3p}{\left(2\pi\right)^3} \frac{1+{\gamma_N}\left(\frac{p}{p_{\rm T}}\right)^4}{1+{\gamma_{\Sigma}}\left(\frac{p}{p_{\rm T}}\right)^4}\right] \label{EEMTf-3}
\te
The spectrum is flat for both regimes $p\ll p_{\rm T}$ and $p_{\rm T}\ll p<p_{\rm L} $, but in the latter the amplitude is diminished by a factor
\be 
\frac{\gamma_N}{\gamma_{\Sigma}}\approx 0.47. \label{EEMTf-4}
\te
We may find the spectrum of entropy fluctuations in a similar way. The entropy density is
\be 
s=\frac{4}3\sigma T_0^3\left[ 1+\frac98{Z_{\rm TT}}^{jk}{Z_{\rm TT}}_{jk} \right]. \label{EEMTf-5}
\te
Therefore
\be 
\left\langle s\right\rangle=\frac{4}3\sigma T^3\left[ 1+\frac9{2\sigma T_0^3}\int\frac{d^3p}{\left(2\pi\right)^3} 
\frac{1+{\gamma_N}\left(\frac{p}{p_{\rm T}}\right)^4}{1+{\gamma_{\Sigma}}\left(\frac{p}{p_{\rm T}}\right)^4}\right] 
\label{EEMTf-6}
\te
Of course, these are the spectra induced by irreducible tensor fluctuations only; the full energy and entropy spectra will also have 
contributions from vector modes.

These nontrivial spectra show that there is a definite redistribution of entropy among the short wavelength modes because 
of nonlinear effects. Although this equilibrium analysis does not allow us to determine the direction of the energy and entropy flows, 
the fact that when equilibrium is reached the short wavelength spectrum is depleted with respect to the linear case suggests the existence of an inverse cascade of entropy \cite{Eyink18b}.

\section{Conclusions and Discussion}\label{concl}
In this paper we began the study of the non-linear hydrodynamics of a real relativistic conformal fluid within the framework of
Divergence Type Theories, which have the advantage that the Second Law of Thermodynamics is satisfied non-perturbatively. 

In Second Order Theories such as DTTs, the fact that non-ideal effects are described by a new independent tensor variable 
permits to enlarge the set of hydrodynamic effects, as now quadrupolar oscillations represented by purely tensor modes are allowed in the flow, besides the scalar and vector ones already present in First Order theories. 

This fact was previously exploited in \cite{nahuel17} to investigate the induction of primordial gravitational waves by the presence 
of these modes in the Early Universe plasma, and also in \cite{calkan16} in the context of Early Universe magnetogenesis.

In this manuscript we began to develop the nonlinear hydrodynamics of real relativistic fluids by studying in a self-consistent way thermally induced tensor fluctuations.

We consider a simple situation where tensor modes are excited by a Gaussian noise with a white spectrum. As was just said, 
this noise is due to the fluid own thermal fluctuations and the spectrum can be computed from the fluctuation-dissipation theorem.

From the analysis in subsection \ref{fdtdtt} this means that there appears an explicitly stochastic source in the conservation equation for the non-equilibrium current $A^{\mu\nu\rho}$, while the energy-momentum conservation equation is unmodified. 
This may be interpreted as if entropy is added to the system,
while keeping constant its energy content. 

Using techniques borrowed from Quantum Field Theory to study non-linear hydrodynamics, such as the Two-Particle-Irreducible Effective
Action and the Martin-Siggia-Rose formalism, we wrote down the evolution equations for the retarded and Hadamard propagators for both the vector and tensor sectors. We first found the lowest linear order expression for the two-point functions and latter non-linear 
fluctuations around equilibrium were considered. It was found that the non-linearities renormalized the relaxation time of the 
theory in a way that induces a depletion of the tensor correlations in the range $p_{\rm T}\ll p\leq p_{\rm L}$ (eq. (\ref{p_t})), with $p_{\rm L}$ the largest value of $p$ for which the loop expansion is consistent. Stated otherwise, we found that tensor fluctuations have a flat spectrum for the largest scales, which turns to a power law $p^{-4}$ spectrum in the small length-scale sector.

In view of the exponentially decreasing time
dependence of the two point functions, these corrections 
to the tensor fluctuation spectrum are significative for 
times shorter than the macroscopic relaxation time,
or else on scales of the order $r < \tau $ which are the 
relevant ones at the initial stages of a heavy-ion collision.

Concerning the entropy to lowest order, the correction to the entropy density is quadratic in the fluctuations and consequently it is also diminished
in the large $p$ range. This result suggests that tensor modes could sustain a turbulent inverse cascade of entropy \cite{Eyink18b}, 
and we intend to study this issue in a forthcoming work.

Besides the studies mentioned just above, other systems where fluid tensor modes can play an important role are   
Neutron Stars \cite{rishke10,FriedSterg13,sterg17} and Early Universe plasmas \cite{nikschlesigl18}, to mention a few. 
In both systems, the fluids are non-ideal relativistic plasma. Therefore it is important to have a solid hydrodynamic theory 
in order to understand the features of those systems. This work is a small step toward that goal and sets the basis for more complete studies of tensor turbulence where energy injection can also be taken into account. 

\appendix

\section{Scaling of the relevant diagrams}\label{app:loopscaling}

In this Appendix we discuss the scaling rules of the subset of diagrams that build 
the different propagators. We perform a general analysis in which we
consider an arbitrary number of loops $L$ and find the condition over a particular combination of parameters that makes the loop 
expansion valid.

It is usual to draw the higher loop corrections in Quantum Field Theory as Feynman diagrams \cite{calhu08,peskschr95}. In any scheme the vertices 
correspond to the interaction action $S_{int}$ and the internal lines to the propagators.

Let us take the noise kernel as an example. Diagrammatically the noise kernel has two external vertices of the same kind:
$\left(Y_{\rm TT}\right.$-$V$-$\left.Z_{\rm T}\right)$ with $Y_{\rm TT}$ in the external lines. In order to simplify, we shall 
consider that in the internal structure of the noise kernel we just have the fields $Y$, $V$ and $Z_{\rm T}$ and vertices of the 
type $\left(Y\right.$-$Z_{\rm T}$-$\left.Z_{\rm T}\right)$ or $\left(Y\right.$-$V$-$\left.V\right)$. In this case the internal 
lines are $\langle VV\rangle$, $\langle VY\rangle$, $\langle VZ_{\rm T}\rangle$, 
$\langle Z_{\rm T}Y\rangle$ and $\langle Z_{\rm T}Z_{\rm T}\rangle$.

Let us call the number of internal vertices $v$ and internal lines $j$ of the different types as $v(YVV)$, $v(YZ_{\rm T}Z_{\rm T})$, 
$j(VV)$, $j(VY)$, $j(VZ_{\rm T})$, $j(Z_{\rm T}Y)$ and $j(Z_{\rm T}Z_{\rm T})$. If the number of loops is $L$ we must have the following 
constraints
\bea
2+2\,v(YVV)&=&2\,j(VV)+j(VY)+j(VZ_{\rm T})\nn
2+2\,v(YZ_{\rm T}Z_{\rm T})&=&2\,j(Z_{\rm T}Z_{\rm T})+j(Z_{\rm T}Y)+j(VZ_{\rm T})\nn
v(YVV)+v(YZ_{\rm T}Z_{\rm T})&=&j(VY)+j(Z_{\rm T}Y)\nn
j(VV)+j(VZ_{\rm T})+j(Z_{\rm T}Z_{\rm T})&=&L+1
\label{lpsclg-1}
\tea
Every component of the structure (internal lines, internal and external vertices and 
loop-integrals) has a scaling factor. The complete diagram scales as the product of the scaling factors of each component. 
From Tables \ref{vectorcausal} and \ref{vectorcorr} we extract the rules for the propagators, namely
\bea
\langle VV\rangle&\to& C_{VV}\sim \left(\sigma T_0^3\right)^{-1}\nn
\langle VY\rangle&\to& C_{VY}\sim \left(\sigma T_0^4\right)^{-1}\nn
\langle VZ_{\rm T}\rangle&\to& C_{VZ_{\rm T}} \sim \left(\sigma T_0^3p\right)^{-1}\nn
\langle Z_{\rm T}Y\rangle&\to& C_{Z_{\rm T}Y} \sim\left(\sigma T_0^4p\right)^{-1}\nn
\langle Z_{\rm T}Z_{\rm T}\rangle&\to& C_{Z_{\rm T}Z_{\rm T}} \sim \left(\sigma T_0^3p^2\right)^{-1}. \label{lpsclg-2}
\tea
From eqs. (\ref{ipdf-22}) and (\ref{ipdf-24}) we get the vertex scaling
\bea
(YVV)&\to&V_{YVV}\sim \sigma T_0^4p\nn
(YZ_{\rm T}Z_{\rm T})&\to&V_{YZ_{\rm T}Z_{\rm T}}\sim \sigma T_0^4p^3. \label{lpsclg-3}
\tea
We must also add a factor of $p^3$ for each loop integral and a factor $1/c_Vp$ for each integral over the time labels attached to the internal vertices. We extract the singular part of the time-dependence, thereby introducing a new overall factor $\delta(t-t')/(c_V p)$. So finally the complete diagram scales as
\bea
\delta(p-p')\delta(t-t')\,\frac{a^2\,T_0^7\,p}{c_V\sigma }\,  \left(\frac{p^3}{c_V^2\sigma  T_0^3}\right)^L.\label{conditionloop}
\tea
If we define $p_{\rm L}^3=c_V^2\sigma  T_0^3$, the condition to make the loop expansion consistent reads $p<p_{\rm L}$.

\acknowledgments
N.M.G. thanks P. Minnini for fruitful discussions. 
The work of E.C. was supported in part by CONICET and Universidad de Buenos Aires. 
N.M.G. is supported by a fellowship of Universidad de Buenos Aires.
A.K. acknowledges financial support from FAPESB grant FAPESB-PVE-015/2015/PET0013/2016, and support from
Universidade Estadual de Santa Cruz.

\end{document}